\documentclass[twocolumn]{emulateapj}
\usepackage[bookmarks,bookmarksopen,colorlinks,linkcolor={blue},
  citecolor={blue},urlcolor={blue}]{hyperref}
\usepackage{amsmath}
\usepackage{natbib}
\usepackage{color}
\usepackage[x11names]{xcolor}
\usepackage{graphics}
\usepackage{enumerate}
\usepackage{setspace}
\newcommand{\vsini}[1]{$v\sin{i_s}$#1}
\newcommand{\mps}[1]{m s$^{-1}$#1}
\newcommand{\nrv}[1]{$N_{\text{RV}}$#1}
\newcommand{\sigK}[1]{$\sigma_{K}$#1}
\newcommand{\sigRV}[1]{$\sigma_{\text{RV}}$#1}
\newcommand{\sigeff}[1]{$\sigma_{\text{eff}}$#1}
\newcommand{\sigact}[1]{$\sigma_{\text{act}}$#1}
\newcommand{\sigplan}[1]{$\sigma_{\text{planets}}$#1}
\newcommand{\sigfloor}[1]{$\sigma_{\text{floor}}$#1}
\newcommand{\sigrp}[1]{$\sigma_{r_p}$#1}

\newcommand{\texp}[1]{$t_{\text{exp}}$#1}
\newcommand{\Rhk}[1]{$\log{R_{\text{HK}}'}$#1}
\newcommand{\prot}[1]{$P_{\text{rot}}$#1}
\defcitealias{sullivan15}{S15}

\newcommand{\cdbox}[1]{%
  \colorlet{currentcolor}{.}%
  {\color{Blue1}%
    \dbox{\color{currentcolor}#1}}%
}
\usepackage{dashbox,framed,color,ocg-p}
\newcommand{\ToggleLayer}[2]{%
  \leavevmode
  \pdfstartlink user {
    /Subtype /Link
    /Border [0 0 0]%
    /A <<
      /S/JavaScript
      /JS (
         var aOCGs = this.getOCGs(), Layer;
         var Layers = "#1".split(","), Active = -1, i, l;
         for (l=0; l<Layers.length; l++) {
           Layer = Layers[l];
           for (i=0; aOCGs && i<aOCGs.length; i++) {
             if (aOCGs[i].state && aOCGs[i].name == Layer) {
               Active = l;
               aOCGs[i].state = false;
             }
           }
           if (Active >= 0) break;
         }
         if (Active == -1) {
           for (l=0; l<Layers.length; l++) {
             if (Layers[l] == "") Active = l;
           }
         }
         Active = Active + 1;
         if (Active == Layers.length) Active = 0;
         Layer = Layers[Active];
         for (i=0; aOCGs && i<aOCGs.length; i++) {
           if (aOCGs[i].name == Layer) aOCGs[i].state = true;
         }
      )
    >>
  }#2%
  \pdfendlink
}

\shortauthors{Cloutier et al.}
\shorttitle{RV follow-up of TESS planets}

\begin{document}
\title{Quantifying the Observational Effort Required for the Radial Velocity Characterization of TESS Planets}
\author{Ryan Cloutier\altaffilmark{1,2,3}}
\author{Ren\'e Doyon\altaffilmark{3}}
\author{Fran\c{c}ois Bouchy\altaffilmark{4}}
\author{Guillaume H\'ebrard\altaffilmark{5}}

\altaffiltext{1}{Dept. of Astronomy \& Astrophysics, University
of Toronto. 50 St. George Street, Toronto, Ontario, M5S 3H4, Canada}
\altaffiltext{2}{Centre for Planetary Sciences,
Dept. of Physical \& Environmental Sciences, University of
Toronto Scarborough. 1265 Military Trail, Toronto, Ontario, M1C 1A4, Canada}
\altaffiltext{3}{Institut de recherche sur les exoplan\`{e}tes,
D\'{e}partement de physique, Universit\'{e} de Montr\'{e}al.
2900 boul. Édouard-Montpetit, Montr\'{e}al, Quebec, H3T 1J4, Canada}
\altaffiltext{4}{Observatoire Astronomique de l'Universit\'{e} de Gen\`{e}ve,
  51 Chemin des Maillettes, 1290 Versoix, Switzerland}
\altaffiltext{5}{Institut d'astrophysique de Paris}

\begin{abstract}
  The Transiting Exoplanet Survey Satellite will conduct a 2-year long wide-field survey searching for
  transiting planets around bright stars. Many TESS discoveries will be amenable to mass
  characterization via ground-based radial velocity measurements with any of a growing suite of
  existing and anticipated velocimeters in the optical and near-infrared. In this study we
  present an analytical formalism to compute the number of radial velocity measurements---and hence the
  total observing time---required to characterize RV planet masses with the inclusion of either a white
  or correlated noise activity model. We use our model to calculate the total observing time required to measure
  all TESS planet masses from the expected TESS planet yield while relying on our current understanding of
  the targeted stars, stellar activity, and populations of unseen planets which inform the expected radial
  velocity precision. We also present specialized calculations applicable to a variety of interesting TESS
  planet subsets including the characterization of 50 planets smaller than 4 Earth radii which is expected
  to take as little as 60 nights of observation. Although, the efficient RV characterization of
  such planets requires a-priori knowledge of the `best' targets which we argue can be identified prior to
  the conclusion of the TESS planet search based on our calculations. 
  Our results highlight the comparable performance of optical and near-IR spectrographs for most 
  planet populations except for Earths and temperate TESS planets which are more efficiently characterized
  in the near-IR. 
  Lastly, we present an online tool to the community to compute the total observing times
  required to detect any transiting planet using a user-defined spectrograph
  (\href{http://maestria.astro.umontreal.ca/rvfc}{\texttt{RVFC}}).
\end{abstract}

\section{Introduction}
NASA's \emph{Transiting Exoplanet Survey Satellite} \citep[TESS; ][]{ricker15} launched in April 2018,
is conducting a wide-field survey over at least a 2-year long period and is expected to discover
approximately 1700 new transiting exoplanet candidates at a 2-minute cadence around nearby stars over
nearly the
entire sky \citep[][hereafter \citetalias{sullivan15}]{sullivan15}. Due to their proximity,
many candidate TESS planetary systems, or TESS objects-of-interest (TOIs),
will be amenable to precision radial velocity (RV) observations using ground-based velocimeters to
establish their planetary nature and to measure the masses of identified planets.
The population of TESS planets to-be discovered are on average systematically closer than the 2342
validated Kepler planets\footnote{According to the NASA Exoplanet Archive accessed on March 18, 2018.}
of which only 243 ($\sim 10$\%) have been characterized with RVs.

The growing number of precision velocimeters---and their variety---is vast and includes both optical
and near-infrared spectrographs 
(APF; \citealt{vogt14},
CARMENES; \citealt{quirrenbach14},
CORALIE;
ESPRESSO; \citealt{pepe10},
EXPRES; \citealt{jurgenson16},
G-CLEF; \citealt{szentgyorgyi16},
GIANO; \citealt{oliva06},
GIARPS; \citealt{claudi16},
HARPS; \citealt{mayor03},
HARPS-N; \citealt{costentino12},
HARPS-3,
HDS; \citealt{noguchi98};
HIRES; \citealt{vogt94},
HPF; \citealt{mahadevan12},
IRD; \citealt{kotani14},
iLocater; \citealt{crepp16},
iSHELL, \citealt{rayner12},
KPF; \citealt{gibson16},
MAROON-X; \citealt{seifahrt16},
MINERVA; \citealt{swift15},
MINERVA-Red; \citealt{sliski17},
NEID; \citealt{allen18},
NIRPS; \citealt{bouchy17},
NRES; \citealt{siverd16},
PARAS; \citealt{chakraborty08},
PARVI,
PEPSI; \citealt{strassmeier15},
PFS; \citealt{crane10},
SALT HRS; \citealt{crause14},
SOPHIE; \citealt{perruchot11},
SPIRou; \citealt{artigau14},
TOU; \citealt{ge16},
Veloce,
WISDOM; \citealt{fzresz16}).
Given the large number of velocimeters that can be used for RV characterization of TESS
planet masses it is useful to understand the observational effort required to do so. That is, how many radial
velocity measurements---and total observing time---are required to detect the masses of the TESS 
planets at a given significance. Furthermore,
it is critical to access which spectrographs are best-suited to the efficient mass characterization of
each transiting planet found with TESS. To address these questions, here we present an analytical formalism
to compute the number of RV measurements required to detect a transiting planet's mass and apply it to the
expected TESS planet yield from \citetalias{sullivan15}.
Combining these calculations with an exposure time calculator provides estimates of the total
observing time required to measure all TESS planet masses and to complete a variety of interesting science
cases that will be addressed by TESS.

This paper is structured as follows: Sect.~\ref{sect:model} describes our model used to compute the
total observing time for all TESS objects-of-interest, Sect.~\ref{sect:accuracy} compares our model to
results from existing RV follow-up campaigns of known transiting planetary systems,
Sect.~\ref{sect:simulation} describes the application of our model to the expected
TESS planet population, and Sect.~\ref{sect:results} reports the results for all TESS planets and for
various science cases. We conclude with a discussion and conclusions in Sect.~\ref{sect:disc}. 
Lastly, in the Appendix~\ref{app:rvfc}
we describe our freely available web-tool that utilizes the model from Sect.~\ref{sect:model}
to calculate the total observing time required to detect any transiting planet with a
user-defined spectrograph.

\section{Modelling the total observing time required to measure a transiting planet's mass} \label{sect:model}
Here we derive equations to calculate the number of RV measurements \nrv{,}
of an arbitrary star---in our case a TESS object-of-interest (TOI)---required to measure the mass of
its transiting planet at a given
detection significance; i.e. with a particular RV semi-amplitude measurement uncertainty \sigK{.}
Together with calculations of the exposure time, \nrv{} can be used to
compute the total observing time required to detect each TESS planet with RVs.

\subsection{Calculating \sigK{} from the Fisher Information} \label{sect:fisher}
Given an RV time-series of a TOI $\mathbf{y}(\mathbf{t})$ taken a times $\mathbf{t}$,
the amount of information present in the data with regards to the value of
the planet's RV semi-amplitude $K$ is contained in the Fisher information. The amount of
information regarding $K$ is the $K$ measurement uncertainty \sigK{,} that is calculated by
evaluating the Fisher information given a model of the observed stellar RV variations due to the
planet.

The Fisher
information matrix $B$ is defined as the Hessian matrix of the lnlikelihood of the data given a model
where the model is parameterized by a set of $n$ parameters $\boldsymbol{\theta}=\{\theta_1,\dots,\theta_n\}$.
Explicitly,

\begin{equation}
  B_{ij} = - \frac{\partial^2 \ln{\mathcal{L}}}{\partial \theta_i \partial \theta_j}
  \label{eq:fisher}
\end{equation}

\noindent where the indices $i,j=1,\dots,n$ and

\begin{equation}
  \begin{split}
  \ln{\mathcal{L}} = -\frac{1}{2} &\left[(\mathbf{y}(\mathbf{t})-\boldsymbol{\mu}(\mathbf{t}))^T C^{-1} \right.
    (\mathbf{y}(\mathbf{t})-\boldsymbol{\mu}(\mathbf{t})) + \ln{\text{det} C} \\ 
    &+ \left.N_{\text{RV}} \ln{2\pi} \right]
  \end{split}
  \label{eq:lnl}
\end{equation}
  
\noindent is the generalized lnlikelihood of $\mathbf{y}(\mathbf{t})$ given a model
$\boldsymbol{\mu}(\mathbf{t})$. The RV time-series and model are each 1D vectors containing
\nrv{} measurements. The matrix $C$ is the \nrv{} $\times$ \nrv{} covariance matrix of the residual
time-series $\mathbf{y}(\mathbf{t})-\boldsymbol{\mu}(\mathbf{t})$. Once constructed,
the Fisher information matrix can be inverted to return a new covariance matrix; the covariance matrix
of the $n$ model parameters and whose diagonal elements are equal to the measurement variances in the
model parameters.

The model of observed stellar RV variations due to a single orbiting planet is a
keplerian solution. Its general form is written as

\begin{equation}
  \begin{split}
  \boldsymbol{\mu}(\mathbf{t},P,T_0,K,e,\omega) = K [&\cos{(\boldsymbol{\nu}(\mathbf{t},P,T_0,e,\omega)+\omega)} \\
    &+ e\cos{\omega}],
  \end{split}
\end{equation}

\noindent in terms of the star's orbital period $P$, time of inferior conjunction $T_0$,
RV semi-amplitude $K$, orbital eccentricity $e$, argument of periapsis $\omega$, and true anomaly
$\boldsymbol{\nu}$. If we assume that the planet's orbit is circular---as was done in \citetalias{sullivan15}---then
our keplerian model reduces to

\begin{equation}
  \boldsymbol{\mu}(\mathbf{t},P,T_0,K) = -K \sin{\boldsymbol{\nu}(\mathbf{t},P,T_0)}
  \label{eq:kep}
\end{equation}

\noindent where the true anomaly can be expressed as $\boldsymbol{\nu}(\mathbf{t},P,T_0) =2\pi(\mathbf{t}-T_0)/P$.

\subsubsection{Calculating \sigK{} with white RV noise} \label{sect:fisherwhite}
Using the keplerian model given in Eq.~\ref{eq:kep} we can derive a simple analytical expression for
\sigK{} in terms of \nrv{} from the Fisher information under a few more simplifying assumptions. 
From the resulting expression one can fix \sigK{} to a desired measurement value and calculate \nrv{}
required to measure $K$ at that precision. As a initial assumption, where we will assume that the observed
RV noise is Gaussian distributed, or white.
This assumption is used in the majority of RV analyses in the literature and
is especially applicable to planets with semi-amplitudes much greater than the measured point-to-point RV rms.
Assuming white noise, the covariance matrix $C$ in Eq.~\ref{eq:lnl} is diagonal with RV measurement variances
$\boldsymbol{\sigma}_{\text{RV}}^2(\mathbf{t})$ along the diagonal. The resulting lnlikelihood reduces to

\begin{equation}
  \ln{\mathcal{L}} = -\frac{1}{2} \sum^{n_{\text{RV}}}_{i=1} \left( \frac{y(t_i)-\mu(t_i)}{\sigma_{\text{RV}}(t_i)} \right)^2,
  \label{eq:lnl2}
\end{equation}

\noindent modulo a constant offset that is independent of the model parameters.

The second simplifying assumption is rather
than considering the full measurement uncertainty time-series $\boldsymbol{\sigma}(\mathbf{t})$, we will
assume that the RV measurement uncertainty is well-characterized over time by a scalar value \sigRV{;} a common
assumption when deriving model parameter uncertainties from time-series observations \cite[e.g.][]{gaudi07, carter08}.
Thirdly, in our keplerian model 
we will assume that the values of $P$ and $T_0$ are known a-priori with absolute certainty
from the planet's TESS transit light curve. Although this is not strictly true, $P$ and $T_0$ are often
measured at high precision---compared to $K$---when multiple transit events are detected.
Thus we can treat $P$ and $T_0$ as constants rather than as model parameters such
that the set of model parameters in our keplerian RV solution reduces to a single value;
$\boldsymbol{\theta}=\{K\}$. The Fisher information matrix then reduces to the scalar value

\begin{align}
  B &= - \frac{\partial^2 \ln{\mathcal{L}}}{\partial K^2}, \\
  &= \frac{1}{\sigma_{\text{RV}}^2} \sum^{N_{\text{RV}}}_{i=1} \sin^2{\nu(t_i,P,T_0)},  \label{eq:fisher2} \\
  &= \frac{N_{\text{RV}}}{2\sigma^2_{\text{RV}}}, \label{eq:fisher3}
\end{align}  

\noindent where in the final step we have assumed 
that the \nrv{} measurements are uniformly sampled
over the planet's orbital phases such that the summation term in Eq.~\ref{eq:fisher2} averages to one
half of \nrv{.}

The inverse of the expression in Eq.~\ref{eq:fisher3} is the K measurement variance or

\begin{equation}
  \sigma_{\text{K}} = \sigma_{\text{RV}} \sqrt{\frac{2}{N_{\text{RV}}}}.
  \label{eq:sigK}
\end{equation}

\noindent This remarkably simple expression for \sigK{} as a function
of the RV measurement uncertainty and number of RV measurements 
can be rearranged to calculate the value of \nrv{} that is
required to detect $K$ of any transiting planet, with a precision of \sigK{,} when the RV noise
can be accurately treated as white.

\subsubsection{Calculating \sigK{} when relaxing the white RV noise approximation} \label{sect:fisherGP}
In deriving Eq.~\ref{eq:sigK} we must assume that the RV time-series noise was Gaussian
distributed. However, numerous analyses of transiting systems have shown that there exist cases in which this is
a poor assumption. Instead the RV residuals---after the removal of planetary models---can be temporally correlated
often owing to the presence of RV signals arising from stellar activity
\citep[e.g.][]{haywood14, grunblatt15, lopezmorales16, cloutier17b, dittmann17}. In
such cases, a correlated `noise'\footnote{Note that we use the term correlated `noise' whereas---if arising
  from temporally correlated stellar activity---then this is a signal rather than noise but it is not the planetary
  signal that we interested in.}
model must be fit simultaneously with the planetary models to account for all suspected
RV signals and any potential correlations between model parameters. One popular choice of
correlated noise activity model is a Gaussian process (GP) regression model with a quasi-periodic covariance kernel
of the form

\begin{equation}
  k(t_i,t_j) = a^2 \exp{\left[ -\frac{(t_i-t_j)^2}{2 \lambda^2}
      -\Gamma^2 \sin^2{\left(\frac{\pi |t_i-t_j|}{P_{\text{GP}}} \right)} \right]},
  \label{eq:kernel}
\end{equation}

\noindent and covariance matrix elements

\begin{equation}
  C_{ij} = k(t_i,t_j) + \delta_{ij} \sqrt{\sigma_{\text{RV}}^2(t_i) + \sigma_{\text{jitter}}^2},
  \label{eq:cov}
\end{equation}

\noindent where $t_i$ is the $i^{\text{th}}$ observation epoch in $\mathbf{t}$ and
$\delta_{ij}$ is the Kronecker delta function. The covariance of the GP model function is parameterized
by five hyperparameters: the amplitude of the correlations $a$, the exponential timescale $\lambda$,
the coherence parameter $\Gamma$, the periodic timescale $P_{\text{GP}}$, and an additive scalar jitter
$\sigma_{\text{jitter}}$.
When attempting to measure the semi-amplitude of a known transiting planet, along with a quasi-periodic GP
activity model, the full set of model parameters becomes
$\boldsymbol{\theta} = \{K, a, \lambda, \Gamma, P_{\text{GP}}, \sigma_{\text{jitter}} \}$. Hence our new Fisher
information matrix will be $6 \times 6$ and takes into account the dependence of the $K$
measurement precision on the remaining hyperparameters. 

The elements of the Fisher information matrix are computed identically as before from Eqs.~\ref{eq:fisher}
and~\ref{eq:lnl} but now using a non-diagonal covariance matrix $C$ (Eq.~\ref{eq:cov}). The derivation
of the elements of $B$ are provided in Appendix~\ref{app:fishergp}. The general effect of computing \sigK{}
from this new Fisher information matrix is to decrease its expected value given a time-series of fixed
\nrv{} compared to the value obtained when using
Eq.~\ref{eq:sigK}. The resulting \nrv{} is typically larger than in the white noise limit.
However unlike computing \sigK{} in the white noise limit, the value of \sigK{} when including a
GP correlated noise activity model is dependent on the values of the model parameters themselves
and on the time-series due to the $t_i-t_j$ terms in the covariance kernel (Eq.~\ref{eq:kernel})
and the appearance of $\mathbf{y}(\mathbf{t})$ and
$\boldsymbol{\sigma}(\mathbf{t})$ in the lnlikelihood such that

\begin{equation}
  B \to B(\mathbf{t}, \mathbf{y}, \boldsymbol{\sigma_{\text{RV}}}, \boldsymbol{\theta}).
\end{equation}

\noindent Therefore estimating \sigK{} from the Fisher
information including correlated noise requires time-series as input and numerical values for all model
parameters in $\boldsymbol{\theta}$.

We note that when including a GP correlated noise activity model, a simple analytical expression
for \sigK{} in terms of $\sigma_{\text{RV}}$ and \nrv{} cannot be derived. In this case \sigK{} must be
derived as a function of \nrv{}
from time-series of varying \nrv{} used to compute the full Fisher
information using Eq.~\ref{eq:fisher} before calculating the covariance matrix of the model parameters
$C'=B^{-1}$ and ultimately \sigK{} from its corresponding diagonal matrix element:

\begin{equation}
  \sigma_{\text{K}} = \sqrt{C_{11}'}.
  \label{eq:sigKGP}
\end{equation}

\subsection{Calculating \sigRV{} for TOIs in the photon-noise limit} \label{sect:sigrv}
Calculating \sigK{} from RV time-series with either white noise or correlated noise is dependent
in-part on the RV measurement uncertainty. Here we estimate the photon-noise limited
RV measurement precision \sigRV{} following the formalism from \cite{bouchy01}. Their formalism
is used to calculate \sigRV{} given the RV information content contained within a stellar spectrum
over a particular wavelength range of interest.

From \cite{bouchy01} the RV measurement precision \sigRV{} is shown to be

\begin{equation}
  \sigma_{\text{RV}} = \frac{c}{Q \cdot \text{S/N}},
  \label{eq:sigrv}
\end{equation}

\noindent where $c$ is the speed of light, $Q$ is known as the quality factor of the spectrum,
and S/N is the signal-to-noise ratio achieved over the full spectral range considered. The
S/N contains contributions from the total number of photoelectrons $N_{\text{e}^-}$ obtained from
the source and a contribution from readout noise that begins to dominate the noise budget for the
faintest TESS stars. Our S/N prescription is

\begin{equation}
  \text{S/N} = \frac{N_{\text{e}^-}}{\sqrt{N_{\text{e}^-} + N_{\text{ron}}^2}}.
\end{equation}
  
\noindent Throughout this study we assume
  a fixed readout noise per pixel of 5 e$^-$ and a 4 pixel PSF sampling in
  each orthogonal direction on the detector. The corresponding readout noise is therefore  
  $N_{\text{ron}} = 20$ e$^1$. The quality factor

\begin{equation}
  Q = \frac{\sqrt{\sum_i{W_i}}}{\sqrt{\sum_i{A_i}}}
\end{equation}

\noindent is calculated from the noise-free stellar spectrum $A_i$---given in photoelectrons
and evaluated at the wavelengths $\lambda_i$---and from the optimum weighting function given
by

\begin{equation}
  W_i = \left( \frac{\lambda_i^2}{A_i} \right) \left( \frac{\partial A_i}{\partial \lambda_i} \right)^2.
\end{equation}

\noindent The quality factor represents the density of the RV information content in the spectrum
$A_i$.

When computing \sigRV{} from Eq.~\ref{eq:sigrv}
we use model stellar spectra from the PHOENIX-ACES library \citep{husser13}.
The spectrum for each TOI is retrieved based on the star's effective temperature $T_{\text{eff}}$ and 
surface gravity \citepalias[see][]{sullivan15}
assuming solar metallicity. The native cgs units of flux density for each model spectrum
are converted to photoelectrons using the photon energy over the wavelength range provided by the
PHOENIX models ($\lambda \in [0.05,5.5]$ \micron{)} and assuming a fixed nominal
instrumental throughput of 5\%.
The spectrum is then segregated into the spectrograph's various spectral bands whose central wavelengths
and spectral coverage are summarized in Table~\ref{table:bands}. In each spectral band we mask
wavelengths at which the telluric transmission is $< 98$\%, where the
spectral telluric absorption model is calculated at an airmass of 1 from Maunakea at $R=100,000$.
The aforementioned model is obtained from the \texttt{TAPAS} web-tool \citep{bertaux14}.
The remaining spectrum in each band that is largely uncontaminated by telluric absorption
is resampled assuming a fixed 3 pixel PSF sampling of each resolution element.
This spectrum is then convolved with a Gaussian kernel whose full width at half maximum is
$\text{FWHM} = \lambda_0 / R$ where $R$ is the spectral resolution of the spectrograph.
After convolving each spectrum with a Gaussian instrumental profile
the spectrum is convolved with the rotation kernel presented in \cite{gray08}
(c.f. Eq. 17.12) that emulates
the effect of rotational broadening for stars with a non-zero projected rotation velocity \vsini{.}
When computing the rotation kernel we adopt a linear limb-darkening coefficient of $\epsilon=0.6$.
For each star we compute \vsini{} from the known stellar radius \citepalias[see][]{sullivan15}, the
inclination of the stellar spin-axis to the line-of-sight $i_s$---drawn from a narrow geometric
distribution centered on $90^{\circ}$---and the stellar rotation period \prot{,} 
that we sample following the methodology described in Sect.~\ref{sect:act}.

\begin{deluxetable}{cccc}
\tabletypesize{\scriptsize}
\tablecaption{Adopted Spectral Bands \label{table:bands}}
\tablewidth{0pt}
\tablehead{Spectral & Central & Effective & Zero-point \\ Band & Wavelength & Band Width & Flux Density \\ & [\micron{]} &  [\micron{]} & [erg/s/cm$^2$/\micron{]}}
\startdata
$U$ & 0.3531 & 0.0657 & $3.678 \times 10^{-5}$ \\
$B$ & 0.4430 & 0.0973 & $6.293 \times 10^{-5}$ \\  
$V$ & 0.5537 & 0.0890 & $3.575 \times 10^{-5}$ \\
$R$ & 0.6940 & 0.2070 & $1.882 \times 10^{-5}$ \\
$I$ & 0.8781 & 0.2316 & $9.329 \times 10^{-6}$ \\
$Y$ & 1.0259 & 0.1084 & $5.949 \times 10^{-6}$ \\
$J$ & 1.2545 & 0.1548 & $2.985 \times 10^{-6}$ \\
$H$ & 1.6310 & 0.2886 & $1.199 \times 10^{-6}$ \\
$K$ & 2.1498 & 0.3209 & $4.442 \times 10^{-7}$
\enddata
\tablecomments{Values presented here were obtained from the \href{http://svo2.cab.inta-csic.es/theory/fps3/index.php?mode=browse}{SVO Filter Profile Service}. Explicitly, values pertaining to any of the $UBVRI$ bands were obtained from the Generic/Johnson filter set whereas the $YJHK$ bands were obtained from the CFHT/Wircam filter set.}
\end{deluxetable}

After the aforementioned convolutions the PHOENIX model's wavelength grid is resampled to a
constant $\delta \lambda = \lambda_0 / R$ whose value is specified at the center of a reference band.
For the optical spectrograph that will be considered in this study we fix the reference band to be
the $V$ band ($\lambda_0 = 0.55$ \micron{)} whereas the reference band is fixed to the $J$ band
($\lambda_0 = 1.25$ \micron{)} with the near-IR spectrograph considered.

\cite{artigau18} compared the value of \sigRV{} derived from stellar spectra---as we do
here---to the value derived from empirical spectra from HARPS, ESPaDOnS, and CRIRES. They find
small discrepancies between these values at optical wavelengths but claim that the RV precision
derived from model spectra can be over-estimated in the near-IR YJ bands by $\sim 2$ and 
under-estimated in the HK bands by $\sim 0.5$. In deriving our own RV measurement precisions
from spectral models in the optical or near-IR, we apply the multiplicative correct factors
derived in \cite{artigau18}. We note that these corrections were derived based on observations
of a single star (i.e. Barnard's star) whereas the correction factors for other star's with
unique effective temperatures and metallicities may differ from those used here.
Despite that, there exists a clear discrepancy between RV measurement precisions derived
from model and empirical spectra. This is at least true in a subset of spectral bands.
Precise disagreements between model and observationally-derived \sigRV{} in various
bands for \emph{all} spectral types is beyond the scope of this paper but may affect the
photon-noise limited RV precision by a factor $\mathcal{O}(2)$---depending on the
spectrograph---as it does for Barnard's star from the $\sim J$ to $K$ band. Note that the
corresponding effect on total observing times will be $\mathcal{O}(<2)$ due to 
the other contributors to the RV uncertainty in addition to the photon-noise limit (see
Sect.~\ref{sect:noise}).

\subsection{Additional sources of RV noise} \label{sect:noise}
Often when searching for transiting planets in radial velocity the residual rms following the
removal of the maximum a-posteriori keplerian planet
solution exceeds the characteristic RV measurement uncertainty.
This implies the existence of additional sources of RV noise.
The effect of these additional noise sources is detrimental to our ability to
precisely measure planet masses if not properly modelled. Therefore this excess noise should be taken into
account when attempting to estimate \sigK{} using either Eq.~\ref{eq:sigK} or~\ref{eq:sigKGP}.
These additional sources of noise may be
attributed to stellar activity arising from dark spots, plages, and/or faculae, whose corresponding RV
signals are modulated by the stellar rotation period and its harmonics, modulo the amplitude of any differential
rotation \citep[e.g.][]{forveille09, bonfils13, delfosse13b}.
Another source of dispersion in observed RVs may be
from additional planets not seen in-transit \citep[e.g.][Bonfils et al. in prep.]{christiansen17, cloutier17b};
an effect that is especially pertinent when searching for small planets whose RV semi-amplitudes are
less than the characteristic RV uncertainty of the time-series \citep[e.g.][]{astudillodefru17a}.

In order to compute \nrv{} for each TOI using our white noise model we will
absorb the aforementioned additional noise sources into an effective RV uncertainty \sigeff{} rather than the
previously assumed measurement uncertainty derived in the photon-noise limit. The effective RV uncertainty is
written as the quadrature sum of the systematic RV noise floor of the spectrograph \sigfloor{}
(see Table~\ref{table:spectrographs}), the photon-noise limited RV precision, and the RV jitter arising from
additional sources of RV noise such as activity and unknown planets: \sigeff{}
$=\sqrt{\sigma_{\text{floor}}^2+\sigma_{\text{RV}}^2+\sigma_{\text{act}}^2 + \sigma_{\text{planets}}^2}$.
Eq.~\ref{eq:sigK} can then be rearranged for \nrv{} in terms of the effective RV uncertainty:

\begin{equation}
  N_{\text{RV}} = 2 \left( \frac{\sigma_{\text{eff}}}{\sigma_{\text{K}}} \right)^2.
  \label{eq:nrv}
\end{equation}

\noindent In empirical time-series with white RV residuals
\sigeff{} can be estimated from the rms of the residual dispersion following the removal of
all modelled planets.
Eq.~\ref{eq:nrv} is not applicable to 
empirical time-series that require a GP correlated noise activity model.

In Sect.~\ref{sect:accuracy} we will compare the results from real RV campaigns to our 
analytic estimates as a test of their validity. 
However unlike in actual RV time-series, the RV jitter rms
resulting from activity and unknown planets is not known a-priori for any of the TOIs in the
\citetalias{sullivan15} synthetic catalog. We therefore need to employ generalized statistical arguments
to estimate the expected RV jitter from activity and unknown planets for each TOI. These estimates are
described in Sects.~\ref{sect:act} and~\ref{sect:planets} and are based on the empirical distributions of
RV jitter from each of these two physical effects.

\subsection{Estimating RV noise due to stellar activity} \label{sect:act}
Here we will consider estimates of the expected RV jitter due to rotationally modulated stellar
activity; \sigact{.} The arguments presented here are intended to be
representative of field stars in the
solar neighbourhood. Firstly, for each TOI we draw a rotation period \prot{} as a function of the stellar
mass from either the \cite{pizzolato03} empirical distribution for FGK dwarfs ($T_{\text{eff}} > 3800$ K) or from
the \cite{newton16a} empirical distribution for M dwarfs ($T_{\text{eff}} \leq 3800$ K). The corresponding
stellar equatorial velocity is calculated using the stellar radius from \citetalias{sullivan15}. The projected
stellar rotation velocity \vsini{} is then calculated after drawing the inclination of the stellar spin--axis
from a geometrical distribution. The value of \vsini{} acts as a first-order estimate of the star's activity
level \citep[e.g.][]{west15, moutou17}.

For active stars
(\prot{} $\lesssim 10$ days), \cite{oshagh17} showed through simultaneous
K2 photometry and HARPS spectroscopy that monotonic correlations exist between the measured RVs and numerous
spectroscopic activity indicators (e.g. \Rhk{,} FWHM, BIS). Meanwhile quiet stars
(\prot{} $\gtrsim 10$ days) appear to lack such strong
correlations with spectroscopic activity indicators yet do correlate strongly with $F_8$, the photometric
flicker or photometric RMS on timescales $<8$ hours \citep{bastien13}. The $F_8$ parameter has been shown to
correlate with asteroseismic stellar surface gravity measurements \citep{bastien13} that itself correlates
with RV jitter \citep{bastien14}. Thus for inactive FGK stars (\prot{} $\geq 10$ days)
we adopt the following temperature dependent
relation from \cite{cegla14} for the expected RV dispersion due to stellar activity:

\begin{equation}
  \sigma_{\text{act}} = 1 \text{ m} \text{s}^{-1} \times
  \begin{cases}
    84.23 F_8 - 3.35, & T_{\text{eff}} \geq 6000 \text{ K}, \\
    18.04 F_8 - 0.98, & T_{\text{eff}} < 6000 \text{ K}.
  \end{cases} \label{eq:cegla}
\end{equation}

\noindent Typical values of \sigact{} used to derive Eq.~\ref{eq:cegla} from \cite{saar03} range from
$\sim 0.5-10$ \mps{} but with a relatively small median value of $\lesssim 2$ \mps{.}
We sample $F_8$ values---measured in parts-per-thousand---from the empirical \emph{Kepler} distribution
that has been corrected for their intrinsic \emph{Kepler} magnitude \citep{bastien13}. After sampling $F_8$ and its
uncertainty we use Eq.~\ref{eq:cegla} to map to the distribution of \sigact{} for the inactive FGK stars in
the sample of planet-hosting TOIs.

For active FGK stars (\prot{} $<10$ days) we revert to the \Rhk{} activity indicator \citep{noyes84} 
whose distribution among nearby field FGK stars
has been well-characterized \citep{henry96, santos00, wright04, hall07, isaacson10, lovis11}. To estimate \sigact{}
for active FGK stars we compute the corresponding \Rhk{} from \prot{} and $B-V$ using the formalism from
\cite{noyes84}. The following formulation from \cite{santos00} is then used to map \Rhk{} $\to$ \sigact{:}

\begin{equation}
  \sigma_{\text{act}} = 1 \text{ m} \text{s}^{-1} \times
  \begin{cases}
    9.2 R_5^{0.75} & \text{for F dwarfs} \\
    7.9 R_5^{0.55} & \text{for G dwarfs} \\
    7.8 R_5^{0.13} & \text{for K dwarfs},
  \end{cases} \label{eq:santos}
\end{equation}  

\noindent where $R_5 = 10^5 R_{\text{HK}}'$.
The rms of the fits in Eq.~\ref{eq:santos} are 0.17, 0.18, and 0.19 dex for FGK stars respectively.
In deriving Eq.~\ref{eq:santos} as a function of spectral type, \cite{santos00} computed
spectral types for each star in their sample based on their CORALIE spectra. However we lack such spectra and
instead define the boundaries between
FGK stars based on $T_{\text{eff}}$ given the limited information available for the TOIs. The assumed ranges
are $T_{\text{eff,F}} \in (6000,7500]$ K, $T_{\text{eff,G}} \in (5200,6000]$ K, and $T_{\text{eff,K}} \in (3800,5200]$ K.
      By computing \Rhk{} and its uncertainty from sampled values of \prot{} and $B-V$, we can map from \Rhk{} to the
      distribution of \sigact{} for the active FGK stars in the sample of planet-hosting TOIs.

Lastly, for M dwarfs there exists a clean relation between \Rhk{} and \prot{} \citep{astudillodefru17b}.
The correlation saturates at a maximum mean value of \Rhk{} $=-4.045$ for rapid rotators with \prot{} $< 10$
days and falls off with rotation period out to the slowest rotating observed M dwarfs with \prot{} $\gtrsim 100$
days. Explicitly, \cite{astudillodefru17b} find the best-fit step-wise powerlaw to the correlation:

\begin{equation}
  \log{R_{\text{HK}}'}  =
  \begin{cases}
    -1.509 \log{P_{\text{rot}}} -2.550, & P_{\text{rot}} > 10 \text{ days} \\
    -4.045, & P_{\text{rot}} \leq 10 \text{ days}.
  \end{cases} \label{eq:astudillo}
\end{equation}

\noindent The dispersion in the relation for slow rotators is characterized by the uncertainty in the slope and
intercept of 0.007 and 0.020 respectively whereas the dispersion in \Rhk{} for rapid rotators is 0.093. Sampled
values of \prot{} for M dwarfs are used to map to \Rhk{} using Eq.~\ref{eq:astudillo} from which \sigact{} values
are estimated using a relation similar to Eq.~\ref{eq:santos} but extrapolated to M dwarfs 
with $T_{\text{eff}} \leq 3800$ K. The adopted coefficient and powerlaw index for M dwarfs (2 \mps{} and 0.1
respectively) were derived from the set of 23
M dwarfs with $2 \lesssim P_{\text{rot}} \lesssim 150$ days (c.f. Fig. 3 \citealt{cloutier18})
whose RV activity rms was characterized with HARPS (X. Delfosse private communication).
Additional empirical data to further
calibrate these models for M dwarfs are part of an on-going study with HARPS
(Delfosse et al. in prep.) and with a subset of active M dwarfs from CARMENES recently reported \citep{tal-or18}.

We note that the
empirical distributions of RV activity used in this study were derived from observations using optical
spectrographs. However, RV activity signals are known to be chromatic as they largely
depend on the temperature contrast between an active region and the stellar photosphere where the contrast
effect is known decrease from the optical to the near-IR
\citep[e.g.][]{martin06,huelamo08,prato08,reiners10,mahmud11}. Meanwhile, Zeeman broadening of spectral
features increases with wavelength \citep{reiners13}. The dominant source for RV activity as a function of wavelength
and spectral type is not yet fully understood \citep{moutou17} and so we choose to remain agnostic and set
the near-IR RV activity equal to that which is derived in the optical.

\subsection{Estimating RV noise due to unseen planets} \label{sect:planets}
The occurrence rates of planets of various sizes around FGKM stars was well studied with the primary
\emph{Kepler} mission \citep[e.g.][]{fressin13, dressing15a}. For example, the cumulative occurrence rate of planets with
radii $r_p \in [0.8,22]$ R$_{\oplus}$ around FGK stars out to 418 days is $\gtrsim 0.87$ planets \citep{fressin13}.
For M dwarfs, small planets with $r_p \in [0.5,4]$ R$_{\oplus}$ with $P \leq 200$ days appear to be more common with
at least 2.5 such planets per
M dwarf. In the TESS simulations of \citetalias{sullivan15} up to one transiting planet is detected although the
multiplicity of each simulated planetary system is reported.
Here we use the number of additional planets around each TOI---along with the known occurrence rates of
planets---to estimate the RV contribution due to these planets whose transits
are not characterized with TESS.

For TOIs with a reported multiplicity $N_p>1$, we sample the radius $r_p$ and orbital period $P$ of the $N_p-1$
additional planets from the \emph{Kepler}-derived occurrence rates from \cite{fressin13} for FGK dwarfs or from
\cite{dressing15a} for M dwarfs. Because at most only one planet is detected in transit for each TOI and the
transit probability $\propto P^{-2/3}$, we draw the orbital periods of additional planets from values greater than
the reported orbital period of the known TESS planet.
In this way the TESS planet is always the innermost planet in the system and therefore most likely to transit.
However, inner non-transiting planets have been detected in known transiting systems
as a result of a potentially small mutual inclination \citep[$\Delta i \sim 1^{\circ}$;][]{cloutier17b}.

When sampling the planet occurrence rates as a function of $P$ and $r_p$, $f(P,r_p)$, a few caveats arise. Firstly,
$f(P,r_p)$ are reported over a coarse grid. Therefore when drawing a planet with a range of potential orbital
periods and radii we sample the exact value of $P$ and $r_p$ each from a uniform distribution bounded by the edges
of that bin. Secondly, due to the poor detection sensitivity to the smallest planets at large orbital periods,
$f(P,r_p)$ is poorly constrained there. To quantify the values of $f(P,r_p)$ in this regime we assume that
$f(P,r_p)$ evolves smoothly such that in bins where $f(P,r_p)$ is poorly constrained, we can average the measured
values in surrounding bins to populate the previously vacant bin. We restrict all pairs of planets in multi-planet
systems to remain Lagrange stable according to the analytic condition from \cite{barnes06} while assuming circular
orbits for all planets. Lastly, the sampled radii for all additional planets are converted to a planetary mass
$m_p$ according to the empirically derived mean mass-radius relations from \cite{weiss13} or \cite{weiss14},

\begin{equation}
  \frac{m_p}{\text{M}_{\oplus}} = 
  \begin{cases}
    0.440 \left(\frac{r_p}{\text{R}_{\oplus}} \right)^3 + 0.614 \left( \frac{r_p}{\text{R}_{\oplus}} \right)^4, & r_p < 1.5 \text{ R}_{\oplus} \\
    2.69 \left( \frac{r_p}{\text{R}_{\oplus}} \right)^{0.93}, & 1.5 \leq r_p/\text{R}_{\oplus} < 4 \\
    \left( 0.56 \left( \frac{r_p}{\text{R}_{\oplus}} \right) \left( \frac{S}{336.5 S_{\oplus}} \right)^{0.03} \right)^{1.89}, & 4 \leq r_p/\text{R}_{\oplus} < 13.7 \\  
    \mathcal{U}(150,2000), & r_p \geq 13.7 \text{ R}_{\oplus}
  \end{cases}
  \label{eq:MR}
\end{equation}

\noindent where $S$ is the irradiance received by the planet. By adopting the mean mass-radius relation this
formalism neglects to reflect the diversity of exoplanet masses for a given planet radius.  

The sinusoidal keplerian solution with unit amplitude has an rms value of $\sqrt{2}/2 \sim 0.707$.
Therefore for each additional planet $i=1,\dots,N_p-1$,
we can calculate $\sigma_{\text{planets},i}=0.707K_i$ where $K_i$ is the planet's semi-amplitude computed
from its sampled $P_i$, $m_{p,i}$, and host stellar mass.
The total value of \sigplan{}
is calculated by the quadrature addition of $\sigma_{\text{planets},i}$ from each additional planet whose
$K_i >$ \sigRV{.} This latest condition is imposed assuming that additional planets not seen in-transit
but whose semi-amplitudes are large compared to the RV measurement precision will be accurately
modelled in the RV analysis and therefore not contribute to the residual RV rms.

\subsection{Exposure time calculator} \label{sect:etc}
Together with estimates of \nrv{,}
the exposure time \texp{} per TOI can be used to calculate the total observing
time required to detect a TESS planet in radial velocity. For a given star, the exposure time required to
achieve a desired S/N per resolution element will depend on the properties of the spectrograph and telescope
used as well as on the star's magnitude in the spectral bands spanned by the spectrograph.
For each TOI in this study we calculate the exposure time that is required to reach a S/N of at least $100$
at the center of a reference band; $V$ ($\lambda=0.55$ \micron{)} or $J$ ($\lambda=1.25$ \micron{)} for optical
and near-IR spectrographs respectively. The zero point flux densities in each spectral band used to
calculate the S/N $=\sqrt{N_{e^-}}$ per resolution element are reported in Table~\ref{table:bands}.

The benefit of integrating longer is only to achieve a photon-noise
limited RV precision less than a few \sigact{,} is one of diminishing returns because the effective RV precision
becomes dominated by activity and cannot be reduced by increasing \texp{.} Although, it is important to note that
increasing \texp{} permits a better sampling for the spectral CCF thus improving one's ability to accurately
characterize the activity and mitigate its effects. In cases for which
the calculated \texp{} results in \sigRV{} $>$ \sigact{} or
$>$ the expected $K$, we claim that the exposure time is under-estimated. To remedy this we
scale-up \texp{} to achieve \sigRV{} $\lesssim$ \sigact{} and $\lesssim$ the expected $K$.
In our exposure time calculator 
we do impose restrictions on the range of \texp{} that can be considered. Specifically, we restrict
\texp{} $\in [10,60]$ minutes. The shortest permissible exposure time is required to help mitigate
the effects of stellar pulsations and surface granulation which evolve on timescales $\lesssim 10$ minutes
\citep{lovis05, dumusque11}.
The upper limit of 60 minutes is applied in order to limit the total observing time dedicated to a single star.
In practice, the majority of stars requiring $>60$ minutes to achieve the target S/N per resolution element and
beat the RV activity rms,
will result in a correspondingly low \sigK{} and will therefore not be amenable to RV 
characterization within a reasonable timespan.
We do note that this upper limit is chosen somewhat arbitrarily and some observers may wish to increase the
maximum exposure time 
to accommodate certain high-value targets such as temperate Earth-like planets within or
near their host star's habitable zone\footnote{There are approximately nine temperate Earth-like planets
  that will be discovered
  with TESS with $T_{\text{eq}} \in [185,300]$ K and $r_p \leq 1.5$ R$_{\oplus}$ \citepalias{sullivan15}.}.

\section{Model Comparison to Observations}  \label{sect:accuracy}
In Sect.~\ref{sect:model} we derived Eq.~\ref{eq:nrv} for the number of RV observations required to measure
$K$ of a transiting planet with a precision of \sigK{} in an RV time-series with white noise and an
effective RV uncertainty \sigeff{.} Similar calculations can be made of \sigK{(}\nrv{)} in the presence of
correlated noise using the formalism discussed in Sect.~\ref{sect:fisherGP} and Eq.~\ref{eq:sigKGP}. 
Here we compare our analytic estimates of \nrv{-}--under the applicable noise condition---to
observational results from existing RV time-series to ensure that our model provides
an accurate approximation to \nrv{} when applied to the TOIs. We consider two sets of RV time-series
of transiting planetary systems featuring either a white or correlated noise model. The latter models being
restricted to a quasi-periodic GP treatment of correlated RV residuals as was assumed in Sect.~\ref{sect:fisherGP}.
All systems considered must also obey the
assumptions imposed when deriving \sigK{.} To recapitulate, those assumptions are:

\begin{enumerate}
\item the planet's orbital solution is well-approximated as circular.
\item The value of the TESS planet's $P$ and $T_0$ are known to ultra-high precision compared to $K$ such
  that correlations between the measured values of $P$, $T_0$, and $K$ are unimportant.
\item The window function of the RV time-series is (approximately) sampled uniformly over the planet's full
  orbital phase.
\item The white RV time-series have a characteristic scalar RV uncertainty equal to the rms of the RV residuals.
\end{enumerate}

\noindent To make the analytic estimates of \nrv{} for each observed planetary system featuring a white
noise model, we use the $1\sigma$ value of \sigK{} for one planet in the
system. By the nature of the Fisher information the value of \sigK{} for each modelled planet in a
multi-planet system should be equal so long as all ephemerides are well-constrained and
planet semi-amplitudes are not correlated. Indeed, the measured planet
semi-amplitudes should be uncorrelated when all planets' are weakly interacting with distinct orbital
periods. For each RV time-series we set the value of \sigeff{} to the
rms of the RV residuals following the removal of all RV signals modelled by the authors. Some authors
report their residual rms values explicitly whilst others treat the stellar activity signal as
white by fitting an additive scalar jitter parameter that we
add in quadrature to the median RV measurement uncertainty of the time-series to estimate \sigeff{.}
The planetary systems with white noise models
considered in this analysis are summarized in Table~\ref{table:compare_white}.

\begin{deluxetable}{cccccc}
\tabletypesize{\scriptsize}
\tablecaption{Summary of RV Observations for Known Transiting Planets with White RV noise\label{table:compare_white}}
\tablewidth{0pt}
\tablehead{Planetary & \sigeff{} & \sigK{} & Actual & Calculated & Ref. \\ System & [\mps{]} & [\mps{]} & \nrv{} & \nrv{}}
\startdata
GJ 1132 & 3.38 & 0.92 & 25 & 27.0 & 1 \\
GJ 1214 & 4.96 & 1.60 & 21 & 19.2 & 2 \\
HAT-P-4 & 10.20 & 3.00 & 23 & 23.1 & 3 \\
HAT-P-8 & 10.23 & 4.20 & 16 & 11.9 & 3 \\
HAT-P-10 & 6.56 & 2.70 & 13 & 11.8 & 3 \\
HAT-P-12 & 5.05 & 1.60 & 23 & 19.9 & 3 \\
HAT-P-18 & 20.30 & 5.20 & 31 & 30.5 & 3 \\
HAT-P-22 & 9.70 & 3.20 & 18 & 18.4 & 3 \\
HAT-P-24 & 13.00 & 3.60 & 24 & 26.1 & 3 \\
HAT-P-26 & 3.60 & 0.98 & 26 & 27.0 & 3 \\
HAT-P-29 & 11.20 & 4.60 & 11 & 11.9 & 3 \\
HAT-P-33 & 66.00 & 17.50 & 26 & 28.4 & 3 \\
HD 97658 & 2.78 & 0.39 & 96 & 101.6 & 4 \\
HD 149026 & 6.18 & 1.40 & 42 & 39.0 & 3 \\
HD 189733 & 15.00 & 6.00 & 12 & 12.5 & 5 \\
HIP 116454 & 2.01 & 0.50 & 33 & 32.3 & 6 \\
Kepler-10 & 3.55 & 0.34 & 220 & 218.0 & 7 \\
Kepler-78 & 2.60 & 0.40 & 79 & 84.5 & 8 \\
Kepler-93\tablenotemark{a} & 1.86 & 0.27 & 86 & 94.9 & 9 \\
TrES-2 & 18.10 & 5.70 & 19 & 20.2 & 3 \\
TrES-4 & 16.90 & 10.00 & 6 & 5.7 & 3 \\
WASP-1 & 7.00 & 3.20 & 10 & 9.6 & 3 \\
WASP-3 & 17.60 & 5.90 & 15 & 17.8 & 3 \\
WASP-4 & 3.25 & 2.30 & 5 & 4.0 & 3 \\
WASP-16 & 2.62 & 1.60 & 4 & 5.36 & 3 \\
WASP-18 & 8.77 & 6.20 & 6 & 4.0 & 3 \\
WASP-19 & 20.35 & 5.00 & 34 & 33.1 & 3 \\
WASP-24 & 4.62 & 3.20 & 4 & 4.2 & 3 \\
WASP-34 & 3.60 & 1.70 & 8 & 9.0 & 3 \\
XO-2N & 19.00 & 8.00 & 10 & 11.3 & 10 \\
XO-5 & 11.20 & 3.00 & 24 & 27.9 & 3
\enddata
\tablenotetext{a}{HARPS-N measurements only.}
\tablecomments{\textbf{References}: (1) \cite{berta15}, (2) \cite{charbonneau09}, (3) \cite{knutson14}, (4) \cite{howard11}, (5) \cite{bouchy05}, (6) \cite{vanderburg15}, (7) \cite{weiss16}, (8) \cite{howard13}, (9) \cite{dressing15b}, (10) \cite{burke07}}
\end{deluxetable}

To compare our formalism in the presence of correlated noise to observed systems we consider five cases analyzed
with a quasi-periodic GP correlated noise activity model. Namely CoRoT-7 \citep{haywood14},
K2-18 \citep{cloutier17b}, Kepler-21 \citep{lopezmorales16}, Kepler-78 \citep{grunblatt15}, and
LHS 1140 \citep{dittmann17}. The calculated value of \nrv{} for these systems is obtained by evaluating
\sigK{} from Eq.~\ref{eq:sigKGP} using each systems' unique RV time-series $\mathbf{t},\mathbf{y},$ and
$\boldsymbol{\sigma}_{\text{RV}}$ from their
respective papers along with the semi-amplitude and GP hyperparameter values plus uncertainties.
The maximum likelihood parameter values are reported in Table~\ref{table:compare_red}.
Because the model parameters are known to a finite precision we Monte-Carlo sample each model parameter
from a Gaussian distribution
whose mean is equal to its best-fit value and standard deviation equal to the parameter's measured $1\sigma$
uncertainty. Evaluating the Fisher information matrix with $10^3$ model parameter draws results in a
distribution of \sigK{} for each planet from which the distribution of \nrv{} can be calculated using
Eq.~\ref{eq:nrv} after \sigeff{} is derived identically to as in the white noise scenario.

\begin{deluxetable*}{cccccccccccc}
\tabletypesize{\scriptsize}
\tablecaption{Summary of RV Observations for Known Transiting Planets with Red RV noise\label{table:compare_red}}
\tablewidth{0pt}
\tablehead{Planetary & $K$ & $a$ & $\lambda$ & $\Gamma$ & $P_{\text{GP}}$ & $\sigma_{\text{jitter}}$ & $\sigma_{\text{eff}}$ & $\sigma_{\text{K}} $ & Actual & Median & Ref. \\
  System & [\mps{]} & [\mps{]} & [days] & & [days] & [\mps{]} & [\mps{]} & [\mps{]} & \nrv{} & Calculated \nrv{}}
\startdata
CoRoT-7 & 3.42 & 7.0 & 20.6 & 1.0 & 23.8 & 3.44 & 3.93 & 0.66 & 71 & $61.4 \pm 2.7$ & 1 \\
K2-18 & 3.18 & 2.8 & 59.1 & 1.2 & 38.6 & 0.25 & 4.59 & 0.75 & 75 & $80.3 \pm 43.7$ & 2 \\
Kepler-21 & 2.12 & 6.7 & 17.0 & 2.4 & 12.6 & 1.98 & 4.22 & 0.66 & 82 & $106.7 \pm 20.8$ & 3\\
Kepler-78 & 1.86 & 5.6 & 18.5 & 2.5 & 13.3 & 1.10 & 1.85 & 0.25 & 109 & $119.4 \pm 15.0$ & 4\\
LHS 1140 & 5.3 & 9.0 & 277.9 & 2.0 & 134.0 & 3.0 & 9.33 & 1.1 & 144 & $265.5 \pm 64.4$ & 5
\enddata
\tablecomments{\textbf{References}: (1) \cite{haywood14}, (2) \cite{cloutier17b}, (3) \cite{lopezmorales16}, (4) \cite{grunblatt15}, (5) \cite{dittmann17}}
\end{deluxetable*}

\begin{figure}
  \centering
  \includegraphics[width=\hsize]{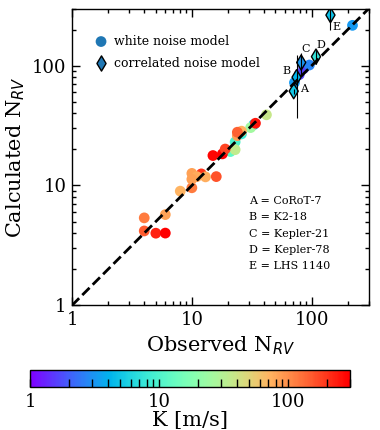}
  \caption{Observed values of \nrv{} compared to the calculated 
    values including either white or correlated RV noise for a suite of known transiting planetary systems.
    The colorbar indicates each planet's RV semi-amplitude. The \emph{dashed line}
    depicts the line $y=x$ wherein calculated \nrv{} correspond exactly to the observed \nrv{.} The region
    above the line depicts where the observed \nrv{}---for a given $K$ measurement
    precision---are less than the value predicted by the Fisher information and would be considered anomalous.
    The region below the line depicts where the observed \nrv{} are larger than the value predicted by the
    Fisher information implying that the $K$ measurement uncertainty may be under-estimated.}
  \label{fig:compare}
\end{figure}

Analytic estimates of \nrv{} are compared to observed values for known planetary
systems in Fig.~\ref{fig:compare}. As evidenced in Fig.~\ref{fig:compare}, the majority of planetary systems
have calculated \nrv{} values in close agreement with observed values for both the white and correlated noise
scenarios. 
This demonstrates that our analytical models for \nrv{} are valid for the majority of cases
with one notable exception. Quantitatively, the rms of the O-C \nrv{} values is 2.6 for the white noise
cases alone and 5.2 for all planetary systems included in Fig.~\ref{fig:compare} with the exception of
the curious outlier LHS 1140.
Our calculated value of \nrv{} for LHS 1140 is overestimated relative to the
size of the RV time-series presented in \cite{dittmann17} from which $K$ is measured to be $5.34 \pm 1.1$
\mps{.} After Monte-Carlo sampling $K$ and the GP hyperparameters from their measurement
uncertainties we calculate a median \nrv{}$=265 \pm 64$ which is $\sim 1.8$ times greater than the actual
time-series size (\nrv{}$=144$) at $1.9\sigma$. The exact cause of this anomalous discrepancy is not known
but may be related to how the GP covariance function is implemented although this investigation is
beyond the scope of this paper. 

\section{Overview of computing \nrv{} for the expected TESS planet population} \label{sect:simulation}
\citetalias{sullivan15} predicted the population of planets that will be discovered with TESS in its
2-minute cadence observing mode. Their results are provided for one realization of their simulations
and contains 1984 TOIs, each with a single transiting planet. The properties of their stellar sample
is copied in Table~\ref{table:stars} for easy reference. This realization contains more detected
planets than the average of their simulations; $\sim 1700$. We treat each TOI in the \citetalias{sullivan15}
sample as a bona-fide exoplanet and not as a false positive. However, some number of TOIs will ultimately
be identified as false positives as historically the false positive rate of transit surveys like Kepler
have yielded higher false positive rates than initially anticipated \citep{sliski14,morton16}.
The properties of the adopted planet population
were derived from planet occurrence rates measured with Kepler circa 2015. Some planetary properties,
particularly the planetary radii, have since been modified slightly following
the reanalysis of Kepler-planet host star properties \citep[e.g.][]{fulton17}.

\begin{figure}
  \centering
  \includegraphics[width=\hsize]{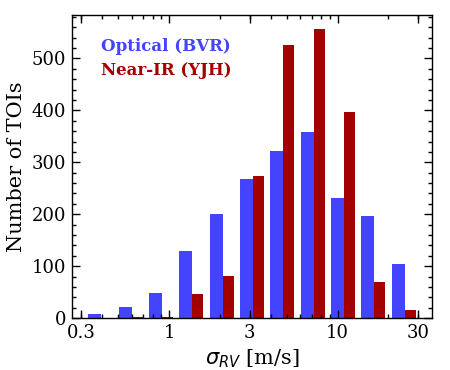}
  \caption{The distributions of the TOI photon-noise limited RV measurement precisions derived from
    PHOENIX stellar models and using the formalism discussed in Sect.~\ref{sect:sigrv}. The values of \sigRV{}
    are computed for each of our fiducial spectrographs in the optical and near-IR with spectral bands $BVR$
    and $YJH$ respectively (see Table~\ref{table:spectrographs}).}
  \label{fig:sigRV}
\end{figure}

Here we compute analytical estimates of \nrv{}
required to measure a planet's RV semi-amplitude---at a given precision---for the entire synthetic catalog
of TESS planets from \citetalias{sullivan15}. Together with the exposure time calculator described in
Sect.~\ref{sect:etc} we calculate the total observing time required to detect each planet. These calculations
require estimates of RV noise sources from the RV noise floor of the employed spectrograph,
photon noise, activity, and additional unseen planets. Photon noise
is dependent on the RV information content contained within the stellar spectrum and varies across spectral
bands. Therefore we consider RV follow-up observations taken with either a fiducial optical spectrograph
or a fiducial near-IR spectrograph. The specifications corresponding to our adopted optical spectrograph are
modelled after the HARPS spectrograph on the 3.6m ESO telescope at La Silla
observatory \citep{mayor03}. The specifications corresponding to our adopted near-IR spectrograph
are modelled after the up-coming NIRPS spectrograph that will join HARPS at the 3.6m ESO
telescope at La Silla observatory in 2019 \citep{bouchy17}. The adopted specifications for our
two fiducial spectrographs are given in Table~\ref{table:spectrographs}. Using the formalism discussed in
Sect.~\ref{sect:sigrv} we calculate \sigRV{} for each TOI with both spectrographs.
For reference, the distributions of TOI \sigRV{} for both spectrographs are shown in Fig.~\ref{fig:sigRV}. 

In Sect.~\ref{app:rvfc}
we will present a web-based \nrv{(}\sigRV{)} calculator that can be used to repeat these calculations but
using \emph{any} spectrograph defined by the user.

\begin{deluxetable}{ccc}
\tabletypesize{\scriptsize}
\tablecaption{Fiducial Spectrograph Specifications \label{table:spectrographs}}
\tablewidth{0pt}
\tablehead{ & Optical & Near-IR}
\startdata
Spectral bands & $BVR$ & $YJH$ \\
Spectral resolution, $R$ & 115,000 & 100,000\tablenotemark{a} \\
Telescope aperture, $D$ [m] & 3.6 & 3.6 \\
RV noise floor, $\sigma_{\text{floor}}$ [\mps{]} & 0.5 & 1.0 \\
Throughput, $\epsilon$ & 0.05 & 0.05
\enddata
\tablenotetext{a}{We consider the higher resolution achievable in the NIRPS High Accuracy Mode relative to its High Efficiency Mode at $R = 75,000$.}
\end{deluxetable}

As discussed in Sects.~\ref{sect:noise},~\ref{sect:act}, and~\ref{sect:planets}, estimates of
\nrv{} are also sensitive to additive RV noise sources such as stellar activity and unseen planets.
However, these astrophysical noise sources are not known a-priori for the \citetalias{sullivan15}
TOIs and therefore must be sampled from known distributions of applicable values of \sigact{}
and \sigplan{.} A Monte-Carlo sampling routine is used for each TOI to sample the aforementioned
quantities from the distributions discussed in Sects.~\ref{sect:act} and~\ref{sect:planets}.
The value of \nrv{} for each TOI also depends on the desired $K$ measurement uncertainty
that is set by the nature of the
follow-up science that one wishes to conduct once the planet's mass has been characterized with RVs.
For example, conventionally the \emph{detection} of a planet's $K$ requires a $3\sigma$ detection
significance; $3 = K/$ \sigK{}. Conversely, targets that will be amenable to atmospheric
characterization via transmission spectroscopy will benefit from a more precise measurement of the
planet's bulk density that in-turn requires a mass detection significance $>3\sigma$.

In the following Sect.~\ref{sect:results} we present results of \nrv{} for the full TESS sample in the
limit of correlated RV noise (see Sect.~\ref{sect:fisherGP}). We also consider four subsamples of TESS
planets each pertaining to a unique science case that will be
addressed by TESS. Each science case merits a unique choice of \sigK{} for a
particular subset of TOIs. Total observing times are then calculated as 
\nrv{}$\cdot (t_{\text{exp}}+t_{\text{overhead}})$. Because the exact value of $t_{\text{overhead}}$ varies
between observatories, we will set $t_{\text{overhead}}=0$ such that its effect can easily
be added to the total observing times later-on for non-zero values. To estimate \nrv{} in the presence
of correlated RV noise we construct time-series of increasing \nrv{} from $10-10^3$ in steps of 90.
For our time-series we adopt a simple uniform window function $\mathbf{t}$ spanning 100 days
which is sufficient to sample the full orbit of $\sim 99$\% of TOIs. Such simplistic
time sampling is admittedly unrealistic given the expected number and frequency of nights
lost due to poor observing conditions (clouds, poor seeing, etc). To that end we
tested more complex window functions 
which included longer baselines and gaps due to observing seasons.
We found results roughly consistent with the uniform window
functions although this need not be true for \emph{any} window function with arbitrary complexity
such as those which are often obtained in practice over many observing seasons.
For each of the 12 time-series with a unique \nrv{} we compute \sigK{} before
interpolating \nrv{(}\sigK{)} to the desired value of \sigK{.} The
initial guesses of the GP hyperparameters $\{\lambda, \Gamma, P_{\text{GP}} \}$ are adopted from
from \cite{dittmann17} (see Methods section \emph{Radial-velocity analysis with Gaussian process regression.})
with the remaining GP hyperparameters set to $a=\sqrt{2}\sigma_{\text{act}}$ and
$\sigma_{\text{jitter}}=\sigma_{\text{planets}}$. The RV time-series $\mathbf{y}(\mathbf{t})$
contains keplerian contribution from the
TESS planet and other sampled planets if applicable, plus correlated noise from a sample of the GP prior
distribution, and white noise featuring contributions from the RV noise floor of the spectrograph  
(Table~\ref{table:spectrographs}) and the photon-noise limited measurement precision:
$\sqrt{\sigma_{\text{floor}}^2 + \sigma_{\text{RV}}^2}$. 
The RV measurement uncertainty time-series $\boldsymbol{\sigma}_{\text{RV}}(\mathbf{t})$
contains the aforementioned value repeated \nrv{} times.

\section{Results for the TESS sample} \label{sect:results}
\subsection{Detecting all TESS planet masses at $3\sigma$} \label{sect:all}
Here we present the results of attempting to detect the masses of \emph{all} TESS planets at $3\sigma$
using either the optical or near-IR spectrograph. As such, we do not make a cut in declination and restrict
targets to half of the sky. Our fiducial spectrographs are intended to be representative of suites of
spectrographs---with comparable on-sky performance---thus providing full sky coverage. 
Realistically not all TESS planets will
be characterized with RVs due to either their small RV semi-amplitude, certain intrinsic stellar
host properties that deter RV observations (e.g. a low apparent
magnitude, rapid projected stellar rotation, or high levels of stellar activity), or simply due to
a lack of available observing time. Despite this fact we present the results for \emph{all} TESS
planets.

Detecting a planet's mass at $3\sigma$ requires a $K$ detection significance that is slightly larger
than three because the calculation of $m_p$ from $K$ is also dependent on other observables such as the
orbital period and stellar mass whose measurement uncertainties contribute to the $m_p$ measurement uncertainty.
To calculate the value of \sigK{}
required to achieve $m_p / \sigma_{\text{m}_\text{p}}=3$ we first assume that the orbital period
of the planet is known to a sufficiently high fractional precision relative to the other parameters
of interest (i.e. $\sigma_{\text{P}}/P \ll 1$) such that its contribution to $\sigma_{m_p}$
can be effectively ignored. Secondly, we assume throughout this study
that all stellar masses are measured with a conservative 
precision of 10\% as many field dwarfs in the solar neighbourhood have their masses measured
with a precision of $\lesssim 10$\% from mass-luminosity relations \citep{delfosse00, torres10}.
However, this assumed fractional precision will not hold for all TOIs as a
subset will have their masses characterized more precisely using other advanced techniques such as
asteroseismology or spectroscopy coupled with precision parallaxes \citep[e.g.][]{vaneylen17,fulton18}.
We note that the calculations presented here represent conservative
values if TOI stellar masses can be determined to a precision higher than 10\%. 
Nevertheless, under our current assumptions a $3.06\sigma$ detection of $K$
(i.e. \sigK{} $=0.327 K$) is required to detect $m_p$ at $3\sigma$.

The median results---over Monte-Carlo realizations---of our calculations
are reported in Table~\ref{table:results} for
each TOI. Specifically, we report the median photon noise-limited RV precision in both the optical and
near-IR spectrographs, \sigact{,} \sigplan{,} \nrv{,} and total observing times in each spectrograph;
$t_{\text{obs,opt}}$ and $t_{\text{obs,nIR}}$. The values of \nrv{---}and the corresponding
$t_{\text{obs}}$---are derived from the general case which includes a GP treatment of correlated RV noise.

\begin{figure}
  \centering
  \includegraphics[width=\hsize]{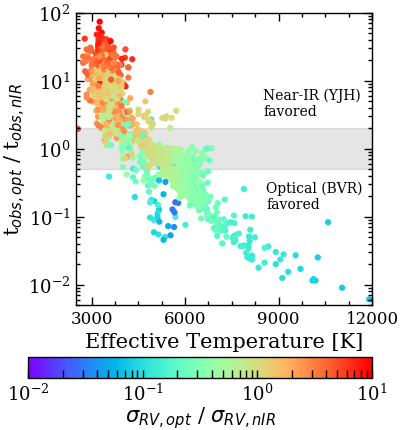}
  \caption{Point estimates of
    the ratio of the median total observing times with an optical spectrograph to with a
    near-IR spectrograph as a function of TOI effective temperature. Colors are indicative of the
    ratio of the photon-noise limited RV precision in the optical to the near-IR. Near-IR RV observations
    are a more efficient means of planet mass characterization when this ratio is a few times unity.
    Conversely, optical RV observations are a more efficient means of planet mass characterization when
    this ratio is a few time smaller than unity. The shaded region spanning the ordinate $\in [1/2,2]$
    approximately depicts where the use of either an optical or near-IR spectrograph offer nearly
    consistent performance.}
  \label{fig:ratio}
\end{figure}

In Fig.~\ref{fig:ratio} we compare $t_{\text{obs,opt}}$ and $t_{\text{obs,nIR}}$ as a function of
TOI effective temperature to ascertain which flavor of spectrograph is favorable for efficient RV planet
mass characterization. A clear trend is discernible with the ratio of the median optical
to the near-IR total observing times decreasing towards earlier spectral types.
Efficient characterization of planets around
late TOIs with $T_{\text{eff}} \lesssim 3800$ K is significantly favoured by the use of near-IR spectrographs
due to the reduced photon noise exhibited by those stars in the near-IR.
Conversely for TOIs with $T_{\text{eff}} \gtrsim 5500$ K, the optical spectrograph is preferred. For
intermediate TOIs the two spectrographs offer nearly consistent performance.

Modulo the effects of rotation and stellar activity the observing time required to detect a transiting
planet with RVs is dependent on the host star's brightness and spectral type which directly effect \sigRV{.} 
Recall that our derived values of \sigRV{} for each TOI are based on stellar spectral models rather than on
empirical spectra such as the M dwarfs observed at high resolution ($R > 80,000$) with the CARMENES visible and
near-IR channels \citep{reiners17}. \cite{reiners17}
claim that the RV information content peaks in the RI bands
between 700-900 nm for all M dwarfs. This empirical evidence somewhat contradicts theoretical calculations
based on model spectra (e.g. \citealt{figueira16}, this study), especially for late M dwarfs whose RV
information content is theorized to peak in the HK bands from 1.4-2.4 $\mu$m. Convergence towards a more
precise scaling of \sigRV{} with wavelength---for stars of various spectral types---will be achieved in the
near future with the onset of multiple new spectrographs from the optical to the near-IR.

\subsubsection{Detecting TESS planet masses versus TOI spectral type}
\begin{figure*}
  \centering
  \includegraphics[width=\hsize]{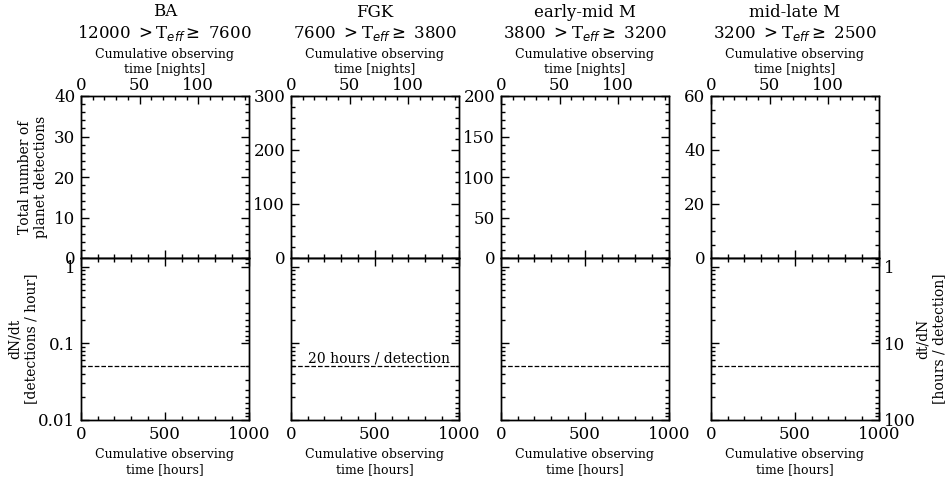}%
  \hspace{-\hsize}%
  \begin{ocg}{fig:optoff}{fig:optoff}{0}%
  \end{ocg}%
  \begin{ocg}{fig:opton}{fig:opton}{1}%
  \includegraphics[width=\hsize]{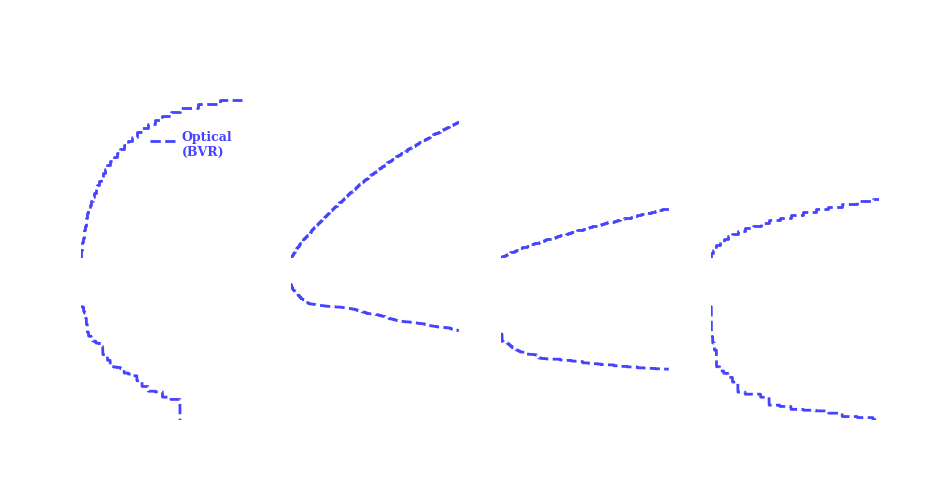}%
  \end{ocg}
  \hspace{-\hsize}%
  \begin{ocg}{fig:niroff}{fig:niroff}{0}%
  \end{ocg}%
  \begin{ocg}{fig:niron}{fig:niron}{1}%
  \includegraphics[width=\hsize]{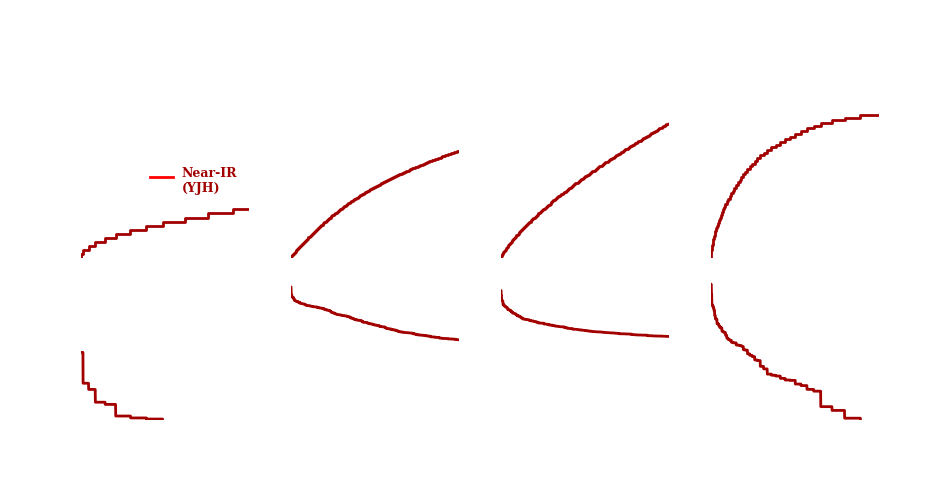}%
  \end{ocg}
  \hspace{-\hsize}%
  \caption{\emph{Upper panel}: the cumulative median observing time to measure the $3\sigma$ RV masses of
    TESS planets as a function of host star spectral type and up to $10^3$ hours. The  
    \ToggleLayer{fig:opton,fig:optoff}{\protect\cdbox{(\emph{dashed blue curves})}} represent the results from
    the optical spectrograph whereas the
    \ToggleLayer{fig:niron,fig:niroff}{\protect\cdbox{(\emph{solid red curves})}} represent the near-IR
    spectrograph. \emph{Lower panel}: the time derivative of the cumulative observing time curves used to indicate
    the RV planet detection efficiency. The \emph{horizontal dashed line} highlights the value of the detection
    efficiency at 20 hours per detection.   
    Note that unlike the lower panels, the upper panels do not share a common ordinate due to the differing
    number of planet detections around stars in each spectral type bin.}
  \label{fig:cumulativeTeff}
\end{figure*}

The cumulative median observing time required to detect TESS planets as a function of TOI spectral type
is shown in Fig.~\ref{fig:cumulativeTeff} up to $10^3$ hours. The results are also given in terms of
the cumulative number of observing nights assuming a notional value of 7 observing hours per night. 
We consider spectral type bins with the following adopted definitions:
mid-late M dwarfs: $2500 \leq T_{\text{eff}}/K < 3200$,
early-mid M dwarfs: $3200 \leq T_{\text{eff}}/K < 3800$,
FGK dwarfs: $3800 \leq T_{\text{eff}}/K < 7600$, and
BA dwarfs: $7600 \leq T_{\text{eff}}/K < 12000$. Spectral type bins are considered separately
because of the clear trend exhibited in total observing times
with either an optical or near-IR spectrograph with $T_{\text{eff}}$ as seen in Fig.~\ref{fig:ratio}.
For example, it is clear that all 39 TESS planets around BA stars can be detected with our 
optical spectrograph in $\sim 140$ nights whereas only $\sim 12$ of those
planets can be detected with the near-IR spectrograph in a thousand hours (i.e. $\sim 143$ nights).
Planet detections around Sun-like
stars (i.e. FGK) are obtained more efficiently with the optical spectrograph with $\sim 251$/964 optical
detections compared to $\sim 198$/964 near-IR detections in a thousand hours.
Efficient M dwarf planet detections favor the near-IR
spectrograph wherein a thousand hours of observing time yields $\sim 165$/927 early-mid M dwarf planets
or nearly all $\sim 54$ mid-late M dwarf planets.
These numbers are reduced to $\sim 60$/927 and $\sim 21$/54 in a thousand hours with our 
optical spectrograph.

Further demonstrated in Fig.~\ref{fig:cumulativeTeff} is the first derivative of the total number of
planet detections with cumulative observing time; $\text{d}N/\text{d}t$.
This quantity describes the efficiency of detecting planets over
time as large values of the derivative highlight when planet masses may be detected in a short amount of observing
time. We will continue by referring to this quantity as the \emph{detection efficiency}. The detection efficiency
can be used to identify after how much total
observing time further planet detections become too observationally expensive. That is that when the detection
efficiency drops below a set threshold value, any additional planet detections will require too much
observing time that may otherwise be spent on potentially more feasible targets.
For reference in Fig.~\ref{fig:cumulativeTeff}
we highlight the value of the inverse time derivative---which is still a measure of detection efficiency---equal
to 20 hours per detection. We suggest this value as a minimum
derivative value. With this threshold value planets around BA stars should be observed for up to $\sim 29$ nights 
with an optical spectrograph before the detection efficiency drops below this threshold value.
Similarly, planets around mid-late M dwarf TOIs should be observed for $\sim 43$ nights
with a near-IR spectrograph . Observing Sun-like and early-mid M dwarf stars can mostly proceed efficiently beyond
$10^3$ hours when using either spectrograph however observing early-mid M dwarfs slowly approaches 20 hours per
detection after $\sim 110$ nights.
Observing all TOIs with the optical spectrograph---or with a network of optical spectrographs of
comparable performance---until we reach a detection efficiency of
20 hours per detection would require $\sim 800$ nights of cumulative observing time.
In that time, $\sim 620$ planets could be detected around TOIs of any spectral type with $V \leq 15$.
Repeating this observing campaign with the near-IR
spectrograph---or with a network of near-IR spectrographs of comparable performance---would require $\sim 1600$ nights
of cumulative observing time. In that time $\sim 1030$ planets could be detected around TOIs
of any spectral type with $J \leq 13.6$.

\subsubsection{Detecting TESS planet masses versus planet type}
\begin{figure*}
  \centering
  \includegraphics[width=\hsize]{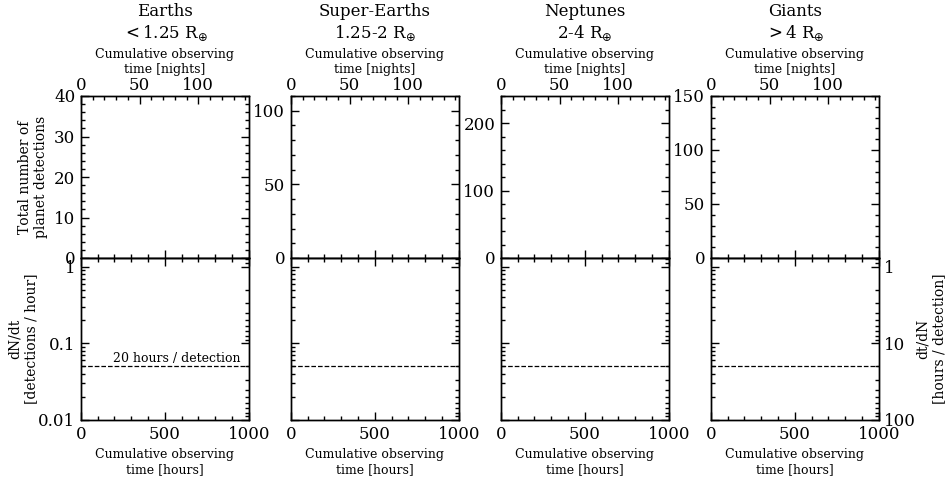}%
  \hspace{-\hsize}%
  \begin{ocg}{fig:optoffr}{fig:optoffr}{0}%
  \end{ocg}%
  \begin{ocg}{fig:optonr}{fig:optonr}{1}%
  \includegraphics[width=\hsize]{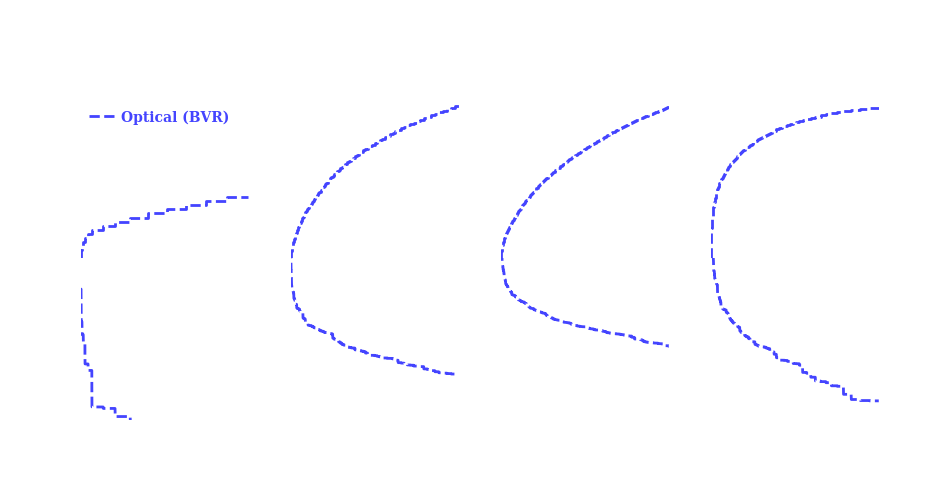}%
  \end{ocg}
  \hspace{-\hsize}%
  \begin{ocg}{fig:niroffr}{fig:niroffr}{0}%
  \end{ocg}%
  \begin{ocg}{fig:nironr}{fig:nironr}{1}%
  \includegraphics[width=\hsize]{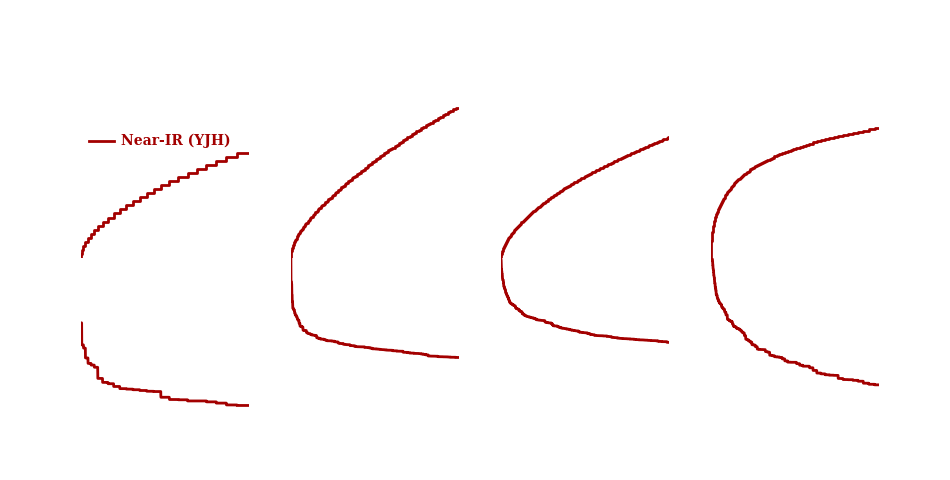}%
  \end{ocg}
  \hspace{-\hsize}%
  \caption{\emph{Upper panel}: the cumulative median observing time to measure the $3\sigma$ RV masses of
    TESS planets as a function of planet type up to $10^3$ hours. Planet type definitions are annotated above
    each column. The  
    \ToggleLayer{fig:optonr,fig:optoffr}{\protect\cdbox{(\emph{dashed blue curves})}} represent the results from
    the optical spectrograph whereas the
    \ToggleLayer{fig:nironr,fig:niroffr}{\protect\cdbox{(\emph{solid red curves})}} represent the near-IR
    spectrograph. \emph{Lower panel}: the time derivative of the cumulative observing time curves used to indicate
    the RV planet detection efficiency. The \emph{horizontal dashed line} highlights the value of the detection
    efficiency at 20 hours per detection.
    Note that unlike the lower panels, the upper panels do not share a common ordinate due to the differing
    number of planet detections around stars in each spectral type bin.}
  \label{fig:cumulativerp}
\end{figure*}

The cumulative median observing times required to detect TESS planets as a function of planet type 
are shown in Fig.~\ref{fig:cumulativerp} up to $10^3$ hours. We consider four types of planet defined by their radii to
be Earths ($<1.25$ R$_{\oplus}$), super-Earths ($1.25-2$ R$_{\oplus}$), Neptunes ($2-4$ R$_{\oplus}$),
and giants ($>4$ R$_{\oplus}$).

Of the 1984 planets in the \citetalias{sullivan15}
TESS sample, 66 are classified as Earths with 26 around stars with
$V \leq \text{median}(V) = 13.5$ and 36 around stars with $J \leq \text{median}(J)=10.7$. Most Earths will
be detected in-transit around M dwarfs ($T_{\text{eff}} \leq 3800$ K) due to their favorable transit depths.
With a thousand hours of total observing time we expect $\sim 26$/66 Earths to be detected with the 
near-IR spectrograph compared to $\sim 15$ detections in the optical. We note that detections of the smallest
planets can be expensive as the detection efficiency exceeds 20 hours per detection
after $\sim 11$ nights or after just $\sim 8$ detections in the near-IR. The detection efficiency
drops more rapidly in the optical to greater than 20 hours per detection after just $\sim 6$ nights or
$\sim 4$ detections.

Super-Earths can be detected rather efficiently in the near-IR to beyond $10^3$ hours and with an optical
detection efficiency better than 20 hours per detection up to $\sim 100$ nights.
We expect to yield $\sim 100$/509 super-Earths with either the optical or near-IR
spectrographs after $10^3$ observing hours.
Neptunes are the most efficiently detected class of planet and with similar detection efficiencies between the two
spectrographs. $\sim 220$/1258 Neptune detections are expected in a thousand hours in the optical
compared to $\sim 180$/1258 in the near-IR. For Neptunes
the detection efficiency remains $<20$ hours per detection for up to 
after $\sim 510$ nights or $\sim 370$ detections in the optical and after $\sim 960$ nights or $\sim 660$ detections
in the near-IR. Lastly, $\sim 130$/151 giant planets are detected in the optical in $\sim 80$ nights at which point the
optical detection efficiency begins to exceed 20 hours per detection. Similarly in the near-IR, $\sim 110$/151 giants planets
are detected in $\sim 85$ nights.

\subsection{Science Case 1: mass characterization of 50 TESS planets with $r_p < 4$ R$_{\oplus}$} \label{sect:lvl1}
The TESS level one science requirement is to measure the masses of 50 small transiting
planets with $r_p < 4$ Earth
radii.\footnote{\href{https://tess.mit.edu/followup/}{https://tess.mit.edu/followup/}}
To date, the vast majority of masses for planets with $r_p <4$ R$_{\oplus}$ have been obtained
with HARPS, HARPS-N, and HIRES. The onset of many up-coming precision velocimeters
will provide many more instruments capable of characterizing
such planets. Similarly to Sect.~\ref{sect:all}, we define a planet mass measurement
requirement for the completion of the TESS level one science requirement of $5\sigma$. 
Given our previous assumptions regarding the measurement precision on $P$ and
$M_s$ (see Sect.~\ref{sect:all}),
a $5\sigma$ mass detection requires a $5.29\sigma$ $K$ detection (i.e. \sigK{} $=0.189 K$).
According to our analytic model for \nrv{} in the white noise limit, a $5\sigma$ mass detection requires
$(0.327/0.189)^2 = 2.99$ more observing time than a $3\sigma$ mass detection.

\begin{figure}
  \centering
  \includegraphics[width=\hsize]{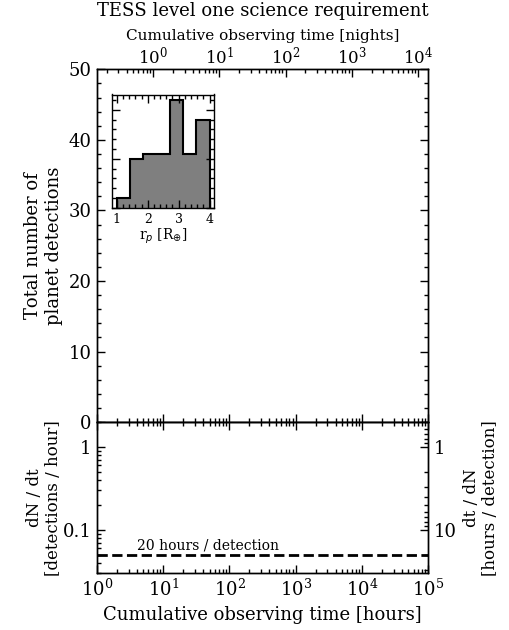}%
  \hspace{-\hsize}%
  \begin{ocg}{fig:optoff50}{fig:optoff50}{0}%
  \end{ocg}%
  \begin{ocg}{fig:opton50}{fig:opton50}{1}%
  \includegraphics[width=\hsize]{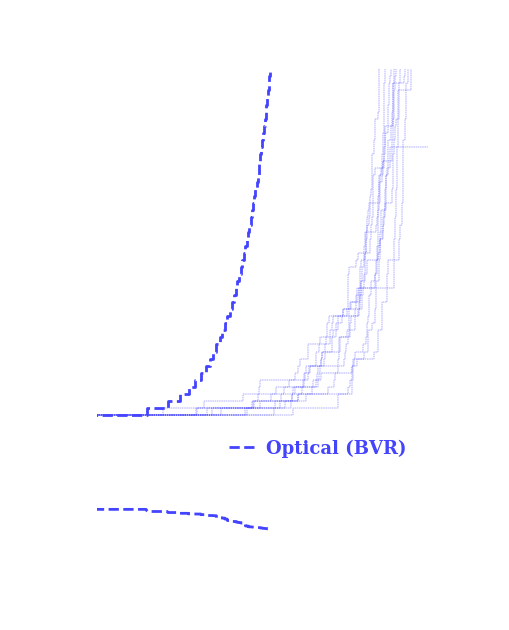}%
  \end{ocg}
  \hspace{-\hsize}%
  \begin{ocg}{fig:niroff50}{fig:niroff50}{0}%
  \end{ocg}%
  \begin{ocg}{fig:niron50}{fig:niron50}{1}%
  \includegraphics[width=\hsize]{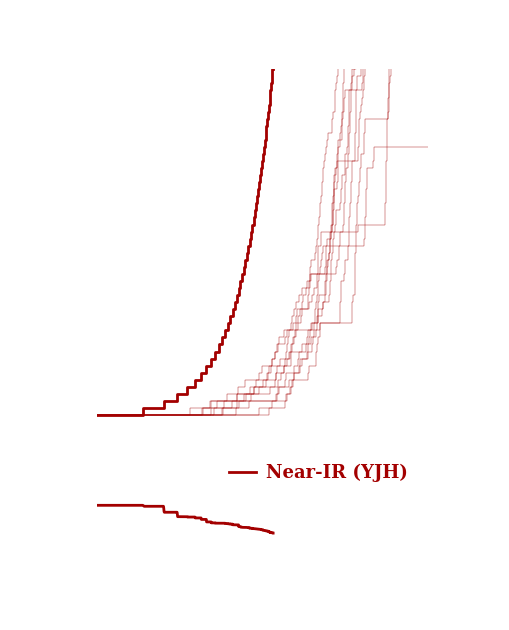}%
  \end{ocg}
  \hspace{-\hsize}%
  \caption{\emph{Top panel}: the cumulative median observing time required to achieve
    the TESS level one science requirement of measuring the masses of 50 planets with
    $r_p <4$ R$_{\oplus}$ at $5\sigma$ with either the optical spectrograph
    \ToggleLayer{fig:opton50,fig:optoff50}{\protect\cdbox{(\emph{dashed blue curves})}} or our
    near-IR spectrograph 
    \ToggleLayer{fig:niron50,fig:niroff50}{\protect\cdbox{(\emph{solid red curves})}}. The set of 
    thin curves are calculated
    from randomly ordered TOI samples whereas the thick curves are calculated from the sorted
    TOIs thus resulting in the most observationally efficient planet detections. The latter curves
    for each spectrograph lie almost exactly on top of each other. \emph{Inset}: a histogram showing the
    joint planet radii distribution of the `best' 50 TESS planets for each spectrograph.
    \emph{Lower panel}: the time derivative of the thick curves shown in the upper panel.
    The value of the detection efficiency equal to 20 hours per detection is highlighted
    by the \emph{horizontal dashed line}.}
  \label{fig:50}
\end{figure}

The cumulative median observing time required to complete the TESS level one science requirement is shown
in Fig.~\ref{fig:50}. Here we calculate the cumulative median observing times from various planet samples:
i) the 50 small TESS planets sorted in ascending order by total observing times
(i.e. the most efficient characterization of 50 small planet masses possible) and ii)
for random subsets of the small TESS planets.
The latter cases correspond to attempting to conduct RV follow-up observations of
\emph{any} subset of small TESS planet up to 50 such planets. In total there are 1833 TESS planets with
$r_p < 4$ R$_{\oplus}$ which causes the cumulative observing time to vary drastically depending on
whether the input planet set is sorted or random. This is evidenced in Fig.~\ref{fig:50} wherein it is
clear that selecting an optimized set---in terms of shortest median observing times---of 50 small TESS planets
is by far the most efficient means of characterizing their masses. Optimized target selection results in the
rapid completion of the TESS level one science requirement in only $\sim 60$ nights with either spectrograph.
The performance of the optical and near-IR spectrographs in completing the TESS level one science
requirement are seen to be comparable when the `best' TOIs are targeted with 31/50 being most efficiently
characterized in the optical and with the remaining 19/50 being done in the near-IR. Selecting
the `best' 50 small planets naturally biases the sample towards larger planets with their larger $K$ values, 
thus making their mass characterization faster with RVs. Despite this, the set of the `best' small TESS planets
with either spectrograph contains 9 super-Earths and 41 Neptunes with the average radius being 2.8 R$_{\oplus}$
and the smallest planet being likely terrestrial at 1.37 R$_{\oplus}$ (c.f. inset of Fig.~\ref{fig:50}).

We note that identifying the `best' small TESS planets cannot be done exactly
until the conclusion of the full TESS planet search. By not focusing on the `best' 50 small
TESS planets and instead opting to obtain RV measurements of any small planet, the total observing time
required to complete the TESS level one science requirement will be longer by more than an order of magnitude
on average with either spectrograph. Although the exact time allotment will depend on the exact
planet sample. However the `best' curves for each spectrograph in Fig.~\ref{fig:50} are nearly
indistinguishable indicating that together optical and near-IR spectrographs will readily complete the TESS
level one science requirement and possibly within weeks of relevant TOIs being announced due to the low number
of required RV measurements, typically \nrv{}$=32$.
Given the efficiency of measuring the `best' small TESS planets at $5\sigma$ ($\sim 5$ hours per
detection), the community may opt to focus on a larger subsample of Earth-like planets or to even increase the
required mass detection significance to $> 5\sigma$ thus enhancing the TESS return of planets smaller than
4 R$_{\oplus}$.

\subsubsection{A-priori estimate of the `best' targets}
Recall that identifying the optimum targets to achieve the TESS level one
science requirement in the most efficient
manner requires observers either to wait until the conclusion of the 2-year long TESS planet search
or to select targets based on a-priori knowledge of the population of the `best' small planets.
The latter scenario is favorable
as it allows targets to be observed with RVs almost concurrently with reported TESS detections thus
leading to the shortest completion time of the TESS level one science requirement; i.e. $\sim 60$
nights or $\sim 400$ hours.
Based on the predicted TESS planet population \citepalias{sullivan15} and the results of our study,
we can predict the properties of the `best' 50 small TESS planets and search for trends that will inform
their selection throughout the actual TESS planet search. 
Here we suggest that identification of these planets can be done in an approximate way
given some combination of intuitively crucial transit observables: the stellar magnitude, $r_p$, and $P$.
The stellar magnitude constrains the photon-noise limited RV measurement precision while the latter two
quantities have a direct effect on $K$ assuming that the mass-radius relation has a positive, non-zero slope
everywhere for planets
smaller than 4 R$_{\oplus}$. After considering numerous combinations of these parameters we find that the
stellar magnitude and the value of the derived-from-transit
quantity $\Omega \equiv r_p^{\alpha}/P^{1/3}$---for some value of $\alpha$---are
good diagnostics for the total observing time required to detect a transiting planet's mass. The
definition of $\Omega$ was selected to resemble the expected RV semi-amplitude $K$ assuming a
positive scaling between $r_p$ and $m_p$ (i.e. $\alpha > 0$) and noting that $K \propto P^{-1/3}$.
In this way, large values of $\Omega$ should correspond to large $K$ values which directly effects the observing
time required to achieve a given mass detection significance as larger signals are more easily detected with
a given Rv measurement precision. We considered various values of $\alpha \in (0,3]$
and found little discrepancy between these values with regards to where in the region of the corresponding
magnitude-$\Omega$ parameter space the `best' 50 small planets sit. Given uncertainties and possible discontinuities
in the mass-radius relation for small planets we opt for $\alpha =1$.

\begin{figure}
  \centering
  \includegraphics[width=\hsize]{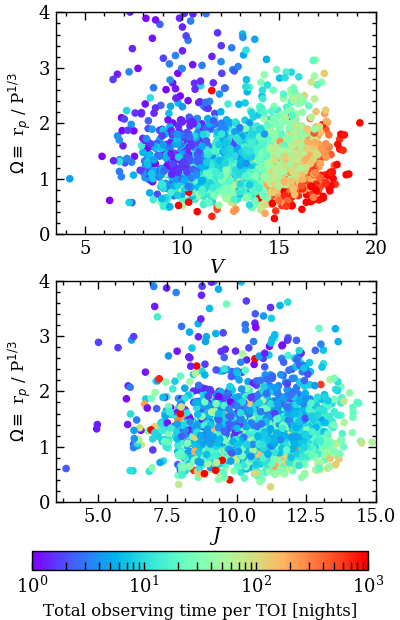}%
  \hspace{-\hsize}%
  \begin{ocg}{fig:bestoff}{fig:bestoff}{0}%
  \end{ocg}%
  \begin{ocg}{fig:beston}{fig:beston}{1}%
  \includegraphics[width=\hsize]{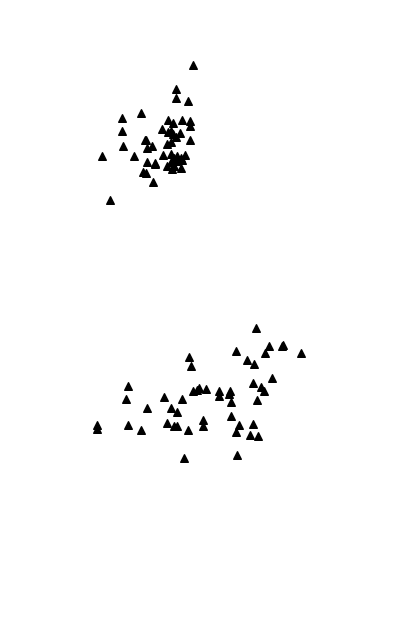}%
  \end{ocg}
  \hspace{-\hsize}%
  \begin{ocg}{fig:regionoff}{fig:regionoff}{0}%
  \end{ocg}%
  \begin{ocg}{fig:regionon}{fig:regionon}{1}%
  \includegraphics[width=\hsize]{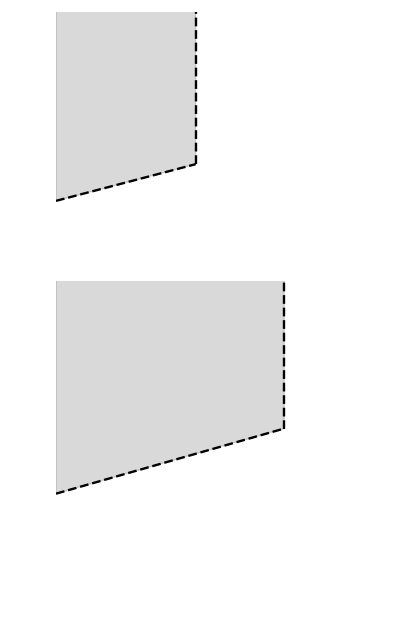}%
  \end{ocg}
  \hspace{-\hsize}%
  \caption{The total observing time per TOI required to detect the planet's mass at $5\sigma$ with
    the optical spectrograph (\emph{upper panel}) and the near-IR spectrograph
    (\emph{lower panel}) as a function of the stellar apparent magnitude and the derived quantity from
    transit observables: $\Omega \equiv r_p/P^{1/3}$.
    The relevant apparent magnitudes for the optical and near-IR spectrographs are $V$ and $J$ respectively. 
    The 50 `best' TOIs with each spectrograph are designated by
    \ToggleLayer{fig:beston,fig:bestoff}{\protect\cdbox{\emph{black diamonds}}}. The
    \ToggleLayer{fig:regionon,fig:regionoff}{\protect\cdbox{\emph{shaded regions}}}
    bounded by the \emph{black dashed lines} approximate the regions of each parameter space with the highest
    likelihood of hosting planets amenable to the most efficient RV mass detections.
    TOIs within the shaded regions and known not to orbit an active host star should be strongly considered
    for rapid RV follow-up campaigns.}
  \label{fig:identify50}
\end{figure}

In Fig.~\ref{fig:identify50} we compare the location of the 50 `best' small planets to the remaining
1833 small planets in the apparent magnitude-$\Omega$ parameter space. For considerations with our
optical and near-IR spectrographs we use the $V$ and $J$ band magnitudes respectively.
We note that in what follows we are marginalizing over stellar rotation and the level of stellar activity, both of
which have a direct effect on our ability to detect planets in RV.

Unsurprisingly, the 50 `best' small TESS planets are localized around bright TOIs and
exhibit an increasing value of $\Omega$ with stellar magnitude. That is, as the TOIs become dimmer,
a larger $\Omega$ is required for a rapid RV mass detection. To encapsulate the region of the parameter space
with the highest likelihood of yielding the most efficient RV planet detections (i.e. the shortest total
observing times), we truncate the outer edge of the
region at a `maximum' apparent magnitude and derive a lower boundary by fitting a linear function to 
$\Omega$ as a function of magnitude for the 
50 `best' small TESS planets before translating the lower boundary downwards to encapsulate 96\%
(i.e. 48 out of 50) of the `best' small TESS planets. The resulting sets of the `best' TOIs are 

\begin{equation}
  \{ \text{TOIs } | V < 10.7, \Omega > 0.09 V + 0.28 \}
  \label{eq:bestopt}
\end{equation}

\noindent for follow-up with our fiducial optical spectrograph and 

\begin{equation}
  \{ \text{TOIs } | J < 11.7, \Omega > 0.14 J - 0.35 \}
  \label{eq:bestnIR}
\end{equation}

\noindent for follow-up with our fiducial near-IR spectrograph. These sets approximately represent the TOIs with
the shortest total observing times and should be seriously considered for rapid RV follow-up observations
if they are known to not orbit a rapid rotator or an overly active star.
Recall that stellar rotation and activity have been marginalized over in the derivation of
Eqs.~\ref{eq:bestopt} and~\ref{eq:bestnIR}.

\subsection{Science Case 2: informing the mass-radius relation of planets across the radius valley} \label{sect:mr}
Accurate characterization of the empirical mass-radius relation for exoplanets
\citep[e.g.][]{weiss13, rogers15, wolfgang16} is an important step towards
understanding the diversity of exoplanet compositions as well as its use as a tool in
predictive studies (e.g. \citetalias{sullivan15}; \citealt{cloutier17a}, \citealt{cloutier18}).
For example, consideration of small
exoplanets ($r_p \leq 4$ R$_{\oplus}$) with masses measured to better than 20\% revealed that a
large fraction of planets with $r_p \lesssim 1.6$ R$_{\oplus}$ are rocky with bulk compositions
consistent with that of the Earth and Venus \citep{dressing15b}.
The transition from bulk rocky compositions to less dense planets with a significant size fraction of
volatile-rich envelope gas has also been shown to occur
between $\sim 1.5-2.5$ R$_{\oplus}$ where a paucity of planets exists \citep{fulton17, vaneylen17}.
The precise characterization of the mass-radius relation in the vicinity of this so-called
\emph{radius valley} will elucidate as to whether or not the valley persists in terms of planet bulk
densities as the peaks in the bi-modal radius distribution are posited to harbour terrestrial and
volatile-rich planets on opposing sides of the radius valley. Characterizing the mass-radius relation
in this regime will greatly benefit from the inclusion of relevant TESS planets.

In order to accurately inform the mass-radius relation of planets across the radius valley 
with TESS planets, we seek a 20\% fractional mass uncertainty (i.e. $5\sigma$ mass detection) following
\cite{dressing15b}. We define TESS planets
of interest as those spanning the radius valley using the period-dependent locus of planet radii---and
its upper and lower bounds---as defined by the powerlaw in \cite{vaneylen17} from asteroseismology.

\begin{figure}
  \centering
  \includegraphics[width=\hsize]{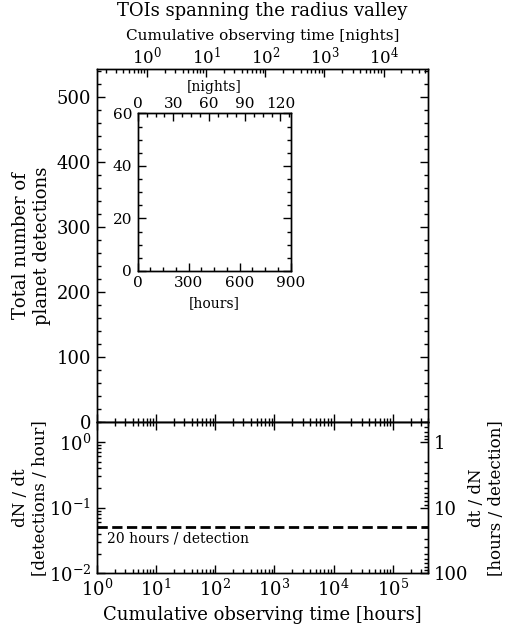}%
  \hspace{-\hsize}%
  \begin{ocg}{fig:optoffMR}{fig:optoffMR}{0}%
  \end{ocg}%
  \begin{ocg}{fig:optonMR}{fig:optonMR}{1}%
  \includegraphics[width=\hsize]{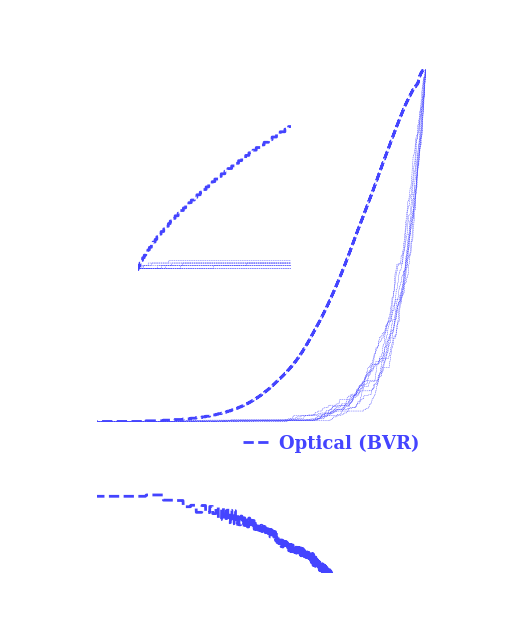}%
  \end{ocg}
  \hspace{-\hsize}%
  \begin{ocg}{fig:niroffMR}{fig:niroffMR}{0}%
  \end{ocg}%
  \begin{ocg}{fig:nironMR}{fig:nironMR}{1}%
  \includegraphics[width=\hsize]{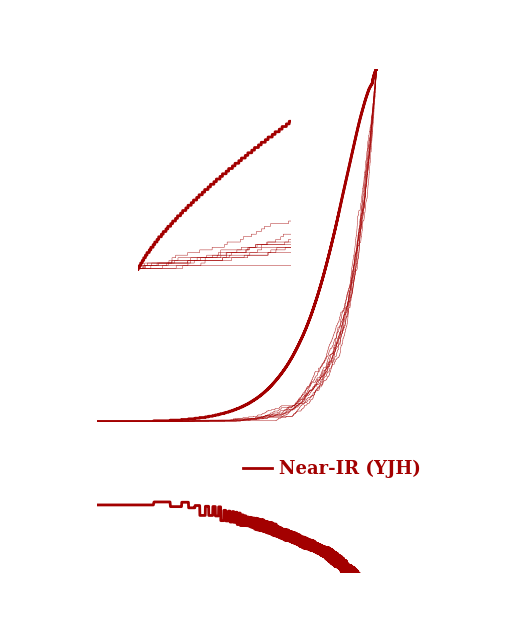}%
  \end{ocg}
  \hspace{-\hsize}%
  \caption{\emph{Top panel}: the cumulative median observing time required to measure the $5\sigma$
    RV masses of
    TESS planets spanning the radius valley ($1.5 \lesssim r_p/\text{R}_{\oplus} \lesssim 2.6$)
    with either the optical spectrograph
    \ToggleLayer{fig:optonMR,fig:optoffMR}{\protect\cdbox{(\emph{dashed blue curves})}} or our
    near-IR spectrograph 
    \ToggleLayer{fig:nironMR,fig:niroffMR}{\protect\cdbox{(\emph{solid red curves})}}. The set of 
    thin curves are calculated from randomly ordered TOI samples whereas the thick curves are
    calculated from the sorted TOIs---with an equal number of planets less than and greater than
    2 R$_{oplus}$---thus resulting in the most observationally efficient planet
    detections. \emph{inset}: focusing on the region up to $900$ cumulative observing hours (i.e.
    $\sim 130$ nights). \emph{Lower panel}: the time derivative of the thick curves shown in the upper panel.
    The value of the detection efficiency equal to 20 hours per detection is highlighted
    by the \emph{horizontal dashed line}.}
  \label{fig:cumulativeMR}
\end{figure}

The cumulative median observing time required to detect relevant TESS planets at $5\sigma$ are
shown in Fig.~\ref{fig:cumulativeMR}. To avoid the bias that the most efficiently observed targets
have towards larger planets and correspondingly larger $K$ on average, the `best' planets in this science
case are selected equally from two bins on either side of the radius valley with $r_p \leq 2$ and
$>2$ R$_{\oplus}$. There are 542 TESS planets that span the radius valley.
RV mass characterization of all such planets will require $\gtrsim 5 \times 10^4$ and $\gtrsim 7000$
observing nights in the optical and near-IR respectively.
Evidently, the cumulative observing time for all TESS planet across the radius valley is
likely too large to complete even with all available spectrographs. Fortunately, not all
542 planets are required to be measured in order to resolve the radius valley in
planet bulk density. If instead we focus on the `best' TESS planets then our detection
efficiency with either spectrograph remains less than 20 hours per detection up to
$\sim 80-130$ nights.
In that time we expect to detect $\sim 55$ planets that span the radius valley with either
spectrograph, if those planets are optimally chosen (c.f. Fig.~\ref{fig:identify50}).
This implies that optical and near-IR spectrographs are equally well-suited to characterizing
the `best' TESS planets across the radius valley with near-IR observations
only becoming more efficient after $\sim 60$ planet detections.
With this sample of TESS planets, the hypothesized rocky/volatile-rich transition can be resolved
and will help in progressing towards potentially resolving the radius/bulk density valley as
a function of host spectral type.

\subsection{Science Case 3: characterization of temperate Earths \& super-Earths}
\begin{figure}
  \centering
  \includegraphics[width=\hsize]{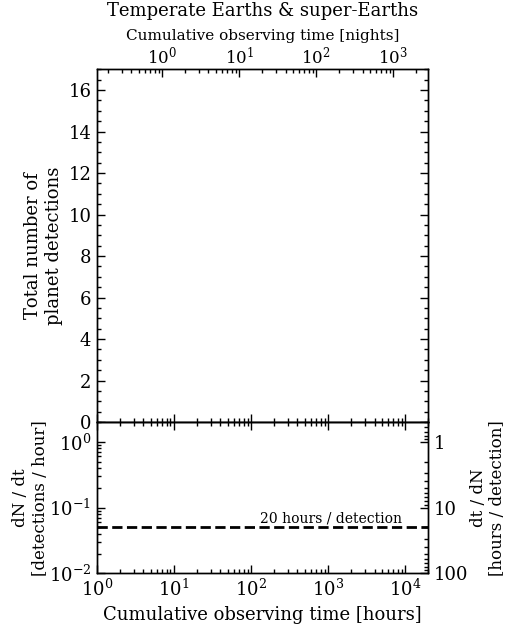}%
  \hspace{-\hsize}%
  \begin{ocg}{fig:optoffHZ}{fig:optoffHZ}{0}%
  \end{ocg}%
  \begin{ocg}{fig:optonHZ}{fig:optonHZ}{1}%
  \includegraphics[width=\hsize]{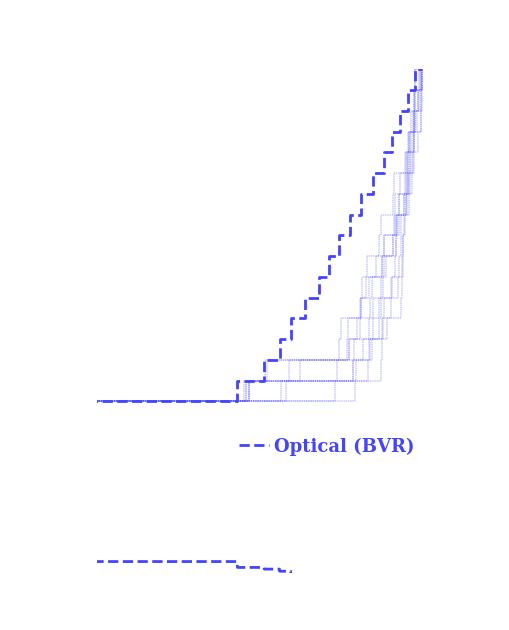}%
  \end{ocg}
  \hspace{-\hsize}%
  \begin{ocg}{fig:niroffHZ}{fig:niroffHZ}{0}%
  \end{ocg}%
  \begin{ocg}{fig:nironHZ}{fig:nironHZ}{1}%
  \includegraphics[width=\hsize]{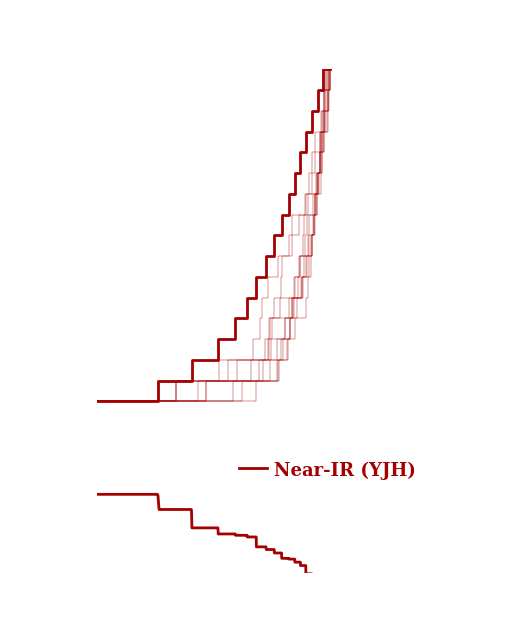}%
  \end{ocg}
  \hspace{-\hsize}%
  \caption{\emph{Top panel}: the cumulative median observing time required to measure the RV masses of
    potentially habitable TESS planets at $3\sigma$ with either the optical spectrograph
    \ToggleLayer{fig:optonHZ,fig:optoffHZ}{\protect\cdbox{(\emph{dashed blue curves})}} or our
    near-IR spectrograph 
    \ToggleLayer{fig:nironHZ,fig:niroffHZ}{\protect\cdbox{(\emph{solid red curves})}}.
    Potentially habitable planets are defined as either Earths or super-Earths with
    $r_p \leq 2 \text{R}_{\oplus}$ and orbit within their host star's habitable zone as
    defined by \citep{kopparapu13}. The set of 
    thin curves are calculated from randomly ordered TOI samples whereas the thick curves are
    calculated from the sorted TOIs thus resulting in the most observationally efficient planet
    detections. \emph{Lower panel}: the time derivative of the thick curves shown in the upper panel.
    The value of the detection efficiency equal to 20 hours per detection is highlighted
    by the \emph{horizontal dashed line}.}
  \label{fig:cumulativeHZ}
\end{figure}

Temperate planets that orbit close to or within their host star's habitable zone are of particular
interest for the search for life.
Here we aim to scrutinize the masses of potentially habitable planets that we define herein as
Earths and super-Earths ($r_p \leq 2$ R$_{\oplus}$) orbiting within the habitable zone (HZ). To define
the HZ we adopt
the `water-loss' and `maximum-greenhouse' HZ limits from \cite{kopparapu13}. There are just 17 such
planets in the \citetalias{sullivan15} synthetic catalog, three of which are smaller than 1.5
R$_{\oplus}$ and are likely to be rocky. 
Fig.~\ref{fig:cumulativeHZ} depicts the cumulative median observing time required to detect the masses
of potentially habitable TESS planets at $3\sigma$.
Due to TESS's limited observational baselines, the majority of HZ TESS
planets---including the full set of 17 potentially habitable TESS planets---orbit M dwarfs and thus
favor near-IR RV follow-up. The cumulative observing time required to measure all potentially
habitable TESS planet masses at $3\sigma$ is $\sim 150$ nights with
the near-IR spectrograph. This is about fifteen times shorter than the total time required
to complete the same task with the optical spectrograph. Regardless of the spectrograph
used, RV follow-up of potentially habitable TESS planets will be an expensive task with
just $\sim 2$ planets detected in $\sim 2.5$ nights after which the detection
efficiency begins to require more than 20 hours per detection.

We caution that with such a small sample of potentially habitable TESS planets that
the numbers presented here regarding the cumulative observing time required to detect such
planets may be misleading. Due to the small number statistics the results
for temperate Earths and super-Earths are highly sensitive to the
true properties of those planetary systems. For example, a potentially
habitable TESS planet may be detected around an M dwarf with $J<10.15$---the brightest TOI with
a potentially habitable planet from \citetalias{sullivan15}---thus resulting in a shorter total
observing time required to detect one such planet with RVs.

\subsection{Science Case 4: characterization of favorable JWST follow-up targets} \label{sect:jwst}
The \emph{James Webb Space Telescope} (JWST) to-be launched in May 2020 will revolutionize our
understanding of transiting exoplanet atmospheres (see \citealt{beichman14} for a summary of science cases).
Many TOIs will represent some of the most interesting targets for atmospheric characterization with
JWST through transmission spectroscopy observations in particular. To quantify the RV requirement needed
to understand the bulk densities of these planets we will consider TOIs that are most amenable to
efficient JWST observations. Specifically, TOIs with their expected S/N of transmission features
$\geq 10$.

For each TOI we calculate the expected S/N in transmission from the expected
differential transmission depth $\Delta D$ of the planet and the photon-noise per spectral bin
$\sigma_{\text{ppm}}$; S/N = $\Delta D / \sigma_{\text{ppm}}$. The value of $\sigma_{\text{ppm}}$
is measured in the $J$-band with spectral resolution $R=50$ (i.e. $\delta \lambda = 25$ nm), an
instrumental throughput of 50\%, and an integration time equal to the planet's full transit duration.
Values of $\Delta D$ for each TESS planet are computed up to five
scale heights in a cloud-free atmosphere using the standard equation

\begin{equation}
  \begin{split}
    \Delta D = & 15 \text{ ppm } \left( \frac{T_p}{250 \text{ K}} \right)
    \left( \frac{\rho}{5.55 \text{ g/cm}^3} \right)^{-1} \\
    &\left( \frac{\mu}{29 \text{ u}} \right)^{-1}
    \left( \frac{R_s}{0.25 \text{ R}_{\odot}} \right)^{-2},
  \end{split}
  \label{eq:transm}
\end{equation}

\noindent where $T_p$ is the planet's isothermal atmospheric temperature (calculated assuming uniform heat
redistribution over the planetary surface and zero albedo), $\rho$ is the planet's
bulk density, $\mu$ is the mean molecular weight of the atmosphere, and $R_s$ is the stellar
radius. The atmospheric mean molecular weights of the TESS planets are not given in \citetalias{sullivan15}
so we adopt a very simplistic prescription of $\mu$ using a step-wise function of H/He-dominated atmospheres
($\mu=2$) to Earth-like atmospheres ($\mu=29$) for planets $\leq 2$ R$_{\oplus}$. However, this simple prescription
is known to be inaccurate but an approximation is necessary to facilitate the exercise of estimating $\Delta D$
for planets yet to be studied in transmission.

According to Eq.~\ref{eq:transm} the interpretation of planetary transmission spectra
heavily relies on a-priori knowledge of the planet's bulk density. The $\rho$ measurement
precision is derived from the measurement precision on both the planet's radius \sigrp{-}--from its
TESS light curve---and on its mass measured from RVs. Due to the cubic dependence
of $\rho$ on $r_p$ compared to its linear dependence on $m_p$, improving a planet's bulk density measurement
precision is most effectively done by reducing \sigrp{} either through more complete
transit data or more precise characterization of the host stellar radius. Because of this, it is not
worthwhile to sit on any TESS planet with RVs to achieve the typical mass detection significance required
to precisely measure the planet's bulk density. For example, given the photometric precision for each TOI
from \citetalias{sullivan15} and a notional stellar radius uncertainty of 10\% \citep{carter08},
achieving a $3\sigma$ bulk
density detection detection would require a $\sim 6.9\sigma$ mass detection on average. Such a precise RV mass
measurement would require a $\sim 7.8\sigma$ $K$ measurement or $(0.327/0.129)^2 = 6.4$ more observing time
than a $3\sigma$ mass detection. We therefore opt for a more reasonable mass detection of $5\sigma$, similarly to
what was pursued when characterizing 50 TESS planets smaller than 4 R$_{\oplus}$ and planets across the radius gap
in Sects.~\ref{sect:lvl1} and ~\ref{sect:mr}.

\begin{figure}
  \centering
  \includegraphics[width=\hsize]{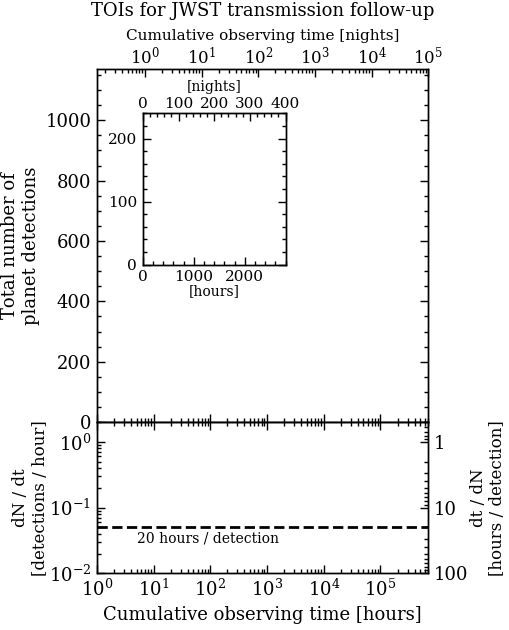}%
  \hspace{-\hsize}%
  \begin{ocg}{fig:optoffjwst}{fig:optoffjwst}{0}%
  \end{ocg}%
  \begin{ocg}{fig:optonjwst}{fig:optonjwst}{1}%
   \includegraphics[width=\hsize]{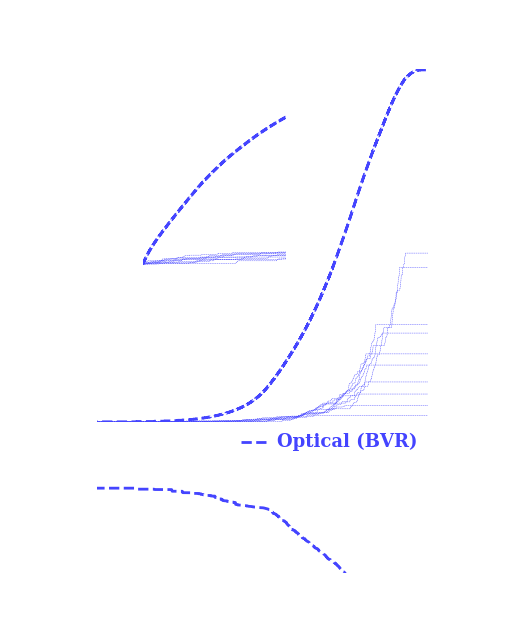}%
  \end{ocg}
  \hspace{-\hsize}%
  \begin{ocg}{fig:niroffjwst}{fig:niroffjwst}{0}%
  \end{ocg}%
  \begin{ocg}{fig:nironjwst}{fig:nironjwst}{1}%
   \includegraphics[width=\hsize]{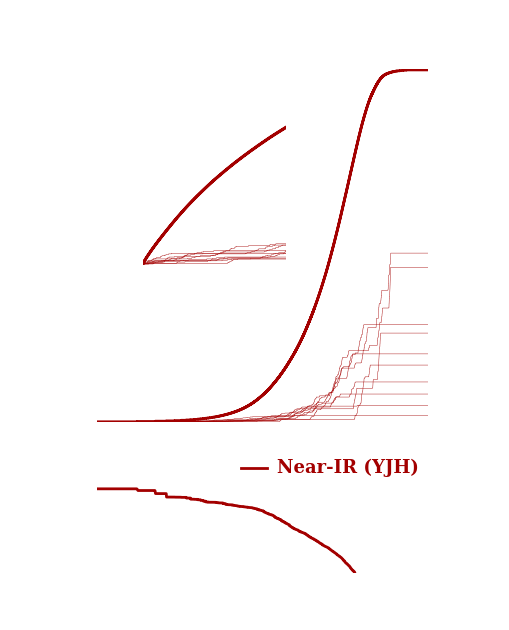}%
  \end{ocg}
  \hspace{-\hsize}%
  \caption{\emph{Top panel}: the cumulative median observing time required to measure the
    RV masses of TESS planets favorable for follow-up transmission spectroscopy observations with
    JWST (i.e. transmission S/N $\geq 10$) at $5\sigma$ with either the optical spectrograph
    \ToggleLayer{fig:optonjwst,fig:optoffjwst}{\protect\cdbox{(\emph{dashed blue curves})}} or our
    near-IR spectrograph 
    \ToggleLayer{fig:nironjwst,fig:niroffjwst}{\protect\cdbox{(\emph{solid red curves})}}. The set of 
    thin curves are calculated from randomly ordered TOI samples whereas the thick curves are
    calculated from the sorted TOIs thus resulting in the most observationally efficient planet
    detections. \emph{inset}: focusing on the region up to $\sim 2800$ cumulative observing hours
    (i.e. $\sim 400$ nights).
    \emph{Lower panel}: the time derivative of the thick curves shown in the upper panel.
    The value of the detection efficiency equal to 20 hours per detection is highlighted
    by the \emph{horizontal dashed line}.}
  \label{fig:trans}
\end{figure}

Fig.~\ref{fig:trans} depicts the cumulative median observing time required to detect the $5\sigma$ masses
of TESS planets that are favorable for JWST follow-up with an expected transmission S/N $\geq 10$.
There are 1169 such TESS planets. By our simple prescription for $\Delta D$ for planets smaller than 2 R$_{\oplus}$
and by imposing a minimum expected transmission S/N $\leq 10$, the sample of 1169 favourable JWST
targets has been restricted to planets larger than 2 R$_{\oplus}$---with a median value
of $r_p = 2.9$ R$_{\oplus}$---due to their systematically larger scale
heights compared to Earths and super-Earths.
Detecting all 1018 Neptunes and all 151 giant planets would require $\sim 10^8$ nights with the near-IR
spectrograph which is $\sim 40$ times
shorter than the total time required using the optical spectrograph.
If follow-up observations are focused on the `best' subset of favorable TESS planets for JWST follow-up,
then the detection efficiency of these planets remains less than 20 hours per detection for up to $\sim 400$
nights wherein $\sim 220$ planets are measured with the near-IR spectrograph. Conversely, the detection
efficiency in the optical drops to 20 hours per detection slightly sooner---after $\sim 360$ nights but with
a similar number of planet detections.
Yet again spectrographs in the optical and near-IR demonstrate a comparable performance
when characterizing the `best' planets favorable for JWST follow-up with some slight improvement in the
optical before its detection efficiency begins to drop off after $\sim 360$ nights. 
It is clear that many interesting TESS planets will be readily
characterized with RVs thus providing a large sample of TESS planets with precisely characterized masses
and bulk densities prior to the launch of JWST.

\section{Discussion and Conclusions} \label{sect:disc}
We have presented calculations of the observing time required to measure the masses of the
expected TESS planet population using ground-based precision radial velocities. Our calculations
are based on analytical estimates (see Sect.~\ref{sect:model})
of the number of RV measurements required to detect a transiting
planet's RV semi-amplitude $K$ at a given precision. When coupled to an exposure time calculator this
yields the total observing time per target. Our main conclusions are summarized below.

\begin{enumerate}
\item The number of RV measurements required to detect a transiting planet's mass is dependent on
  the desired $K$ measurement precision, the rms of the RVs observations (this includes contributions
  from photon-noise, stellar activity, additional unseen planets, and systematic effects),
  and whether or not the residual RV noise is correlated or not.
  Eq.~\ref{eq:nrv} can be used to calculate the number of required RV measurements
  if the RV residuals are uncorrelated,
  otherwise the formalism presented in Sect.~\ref{sect:fisherGP} must be used.
\item Efficient characterization of transiting planet masses for a given planet type (e.g. super-Earths)
  favors targets with small photon-noise limited RV measurement precisions. High precision measurements
  are most readily achieved with optical spectral coverage (i.e. BVR bands in this study)
  for Sun-like stars with $T_{\text{eff}} \gtrsim 5500$ K whereas  M dwarfs with
  $T_{\text{eff}} \lesssim 3800$ K are favored by near-IR spectrographs (i.e. YJH bands in this study).
\item Overall, the relative merits of obtaining precise RVs in the optical compared to in the near-IR are    
  nearly equivalent. That is that RV campaigns aiming to characterize TESS planets can largely be done as
  effectively in either wavelength domain with the exception of the characterization of Earths 
  ($r_p < 1.25$ R$_{\oplus}$) and temperate TESS planets that are preferentially found around M dwarfs.
\item Not all TESS planets will be amenable to RV follow-up observations and selecting random TOIs
  for follow-up is an incredibly inefficient method of target selection. Instead, targets should be
  selected based on their apparent magnitude---to minimize the photon-noise limited RV precision---and
  the transit-derived quantity $\Omega = r_p / P^{1/3}$. The subset of TOIs belonging to either optimal set defined
  by Eqs.~\ref{eq:bestopt} and~\ref{eq:bestnIR} should be strongly considered for immediate RV follow-up.
\item The TESS level one science requirement of measuring the masses of 50 planets with
  $r_P<4$ R$_{\oplus}$ at the 20\% level (i.e. $5\sigma$ mass detection) can be achieved in as little
  as $\sim 400$ hours or $\sim 60$ nights of observation.
\item $\sim 55$ TESS planets spanning the radius gap for small planets
  (i.e. $1.5 \lesssim r_P/\text{R}_{\oplus} \lesssim 2.6$) can be detected efficiently at $5\sigma$
  in $\sim 130$ nights before the detection efficiency drops below 0.05 detections per hour
  (i.e. 20 hours per detection).
\item Only $\sim 2$ temperate super-Earths can be detected efficiently at $3\sigma$ in $\sim 2.5$ nights
  before the detection efficiency exceeds 20 hours per detection.
\item $\sim 220$ Neptunes and giant planets
  amenable to transmission spectroscopy follow-up observations with
  JWST can be detected efficiently with a $5\sigma$ mass detection in $\sim 360-400$ nights before
  the detection efficiency exceeds 20 hours per detection.
\item An online version of the \texttt{Radial Velocity Follow-up Calculator} used throughput this paper on the
  expected TESS planet population is available at
  \url{http://maestria.astro.umontreal.ca/rvfc}. This general-usage tool can be used to calculate the
  number of RV measurements and total observing time required to detect the RV semi-amplitude of any
  transiting planet to a user-defined detection significance and with a user-defined spectrograph.
\end{enumerate}

The results of this paper have been based on the synthetic planet population presented in
\citetalias{sullivan15}. In their study they reported the expected TESS planet population (i.e.
$\sim 1700$ planets) recovered from the TESS 2 minute cadence observing mode of the brightest targets.
Additionally, deeper full frame images at a 30 minute cadence will be released and result in even more planet
candidates, some of which may still be amenable to RV follow-up observations. Another population of TOIs
not considered in our calculations is the population of targets featuring only one or two transit-like events.
If confirmed, the corresponding planets will be interesting in their own right as their orbits will
have systematically longer periods making them cooler and of interest for future habitability studies. 
Lastly, the TESS mission has the possibly of being extended beyond its nominal 2-year long primary mission.
If extended, the extension of the TESS observational baselines will improve the measured ephemerides of confirmed TESS
planets, shed light on the nature of systems exhibiting single transit-like events, and expand the
population of planets discovered with TESS \cite[see the overview in][]{bouma17}.

Further caveats to the planet population from \citetalias{sullivan15} used throughout this study were
addressed by a variety of studies and corresponding updates to the expected TESS planet population
\citep[e.g.][]{ballard18,barclay18}. 
For example, \cite{barclay18} updated
the calculations of \citetalias{sullivan15} by using the TESS Candidate Target List \citep{stassun17}
and consequently updating the number of TESS discoveries including a decreased number of Earths and
super-Earths which will have important implications for the corresponding science cases such as
the bulk density characterization of TESS planets near the radius valley ($r_p \sim 2$ R$_{\oplus}$).
Furthermore, \cite{ballard18} predicted that the planet population around M1-M4 dwarfs was
underestimated. The results imply a larger cumulative planet yield around M dwarfs including many
systems with multiple transiting planets. This has important implications for RV follow-up campaigns
of TOIs as M dwarf planets have the potential to be the most efficiently detected planets in radial velocity.
In practice this will depend on the on-sky performance of many of the up-coming generation of near-IR
velocimeters which have favorable RV measurement precision when observing M dwarfs.
Many up-coming near-IR spectrographs are anticipated to operate with an RV noise floor of $\sim 1$
\mps{} comparable to many high performance optical spectrographs but this may prove
challenging as demonstrated by the CARMENES near-IR channel that seems to yield lower precision
($\sim 2-3$ \mps{)} on early to mid-M dwarfs compared to the sub-1 \mps{} performance achieved in the
optical \citep[][c.f. Fig 6]{reiners17}.
The focus on M dwarf planets with either class of spectrograph is well warranted as these planets
represent some of the most interesting planets in terms of potential habitability and for the
prospect transmission spectroscopy follow-up with JWST.

In Sect.~\ref{sect:jwst} we presented the results for the most favorable TESS targets for
transmission spectroscopy follow-up with JWST. In addition to identifying the `best' such planets
based on the time required to measure their masses (i.e. bulk densities) with RVs,
one may also consider metrics
describing the ease of detecting atmospheric features in either transmission or thermal emission
as in \cite{kempton18}. Favorable TESS planets based on our calculations of
the observing time required to detect their masses in RVs, that also overlap with favorable TESS planets
based on the metrics from \cite{kempton18}, should be strongly considered for rapid RV follow-up.
Similar target selection may be done based on the simulated transmission spectra from \cite{louie18}
and the resulting S/N in transmission.

It is worth reiterating that although TOIs will be frequently reported following the launch and
commissioning of TESS, RV follow-up teams should refrain from targeting just any TOI. Many detected
planets will orbit stars either too dim or too active for efficient RV characterization. We emphasize
that TOIs amenable to RV follow-up can be approximately identified---in real time---if they belong
to one of the sets defined in Eqs.~\ref{eq:bestopt} or~\ref{eq:bestnIR}, or equivalently, if they lie
in one of the shaded regions of either panel in Fig.~\ref{fig:identify50}. However the sets of the
`best' TOIs are derived by marginalizing over the population of additional unseen planets in the
system and the star's intrinsic RV activity. The former source of RV signals will not be illuminated
unless RV follow-up of the system commences but the level of stellar activity can be estimated
from the star's photometric variability in its TESS light curve. In addition to selecting targets
based on Eqs.~\ref{eq:bestopt} and~\ref{eq:bestnIR}, stars with high amplitudes of photometric
variability or jitter should not be considered for efficient RV follow-up campaigns.

\acknowledgements
RC thanks Rapha\"elle Haywood and Jason Dittmann for useful discussions of the GP implementation,
particularly for the LHS 1140 system. RC thanks the Canadian Institute for Theoretical Astrophysics
for use of the Sunnyvale computing cluster throughout this work.
RC is partially supported in this work by the National Science and Engineering Research
Council of Canada.

\appendix
\section{Radial Velocity Follow-up Calculator} \label{app:rvfc}
Our models for the number of RV measurements \nrv{} required to detect a transiting planet
at a given
significance---in the presence of white or correlated RV noise---are generalizable to the majority
transiting planets observed with any velocimeter. Recall that our models are only applicable to
planets on nearly circular orbits, with known ephemerides, and 
whose orbital phase curves will be (approximately) uniformly sampled in the white noise case.
Furthermore, because the results presented throughout this paper
have been regarding a hypothetical planet population and with only two fiducial spectrographs,
we present to the community an online web-tool version of the generalized
\emph{Radial Velocity Follow-up Calculator}\footnote{\url{http://maestria.astro.umontreal.ca/rvfc}}
(\texttt{RVFC}) used throughout this study.

The \texttt{RVFC} is intended to serve the community by providing rapid calculations of
\nrv{} and total observing times for an arbitrary user-defined transiting planet with a user-defined
spectrograph, given parameters of the planet from its
transit light curve, stellar parameters, parameters of the employed spectrograph, RV noise
parameters, and a small number of additional simulation parameters.
The exact input parameters required by the calculator will depend on which of two
possible primary modes-of-operation the user selects. In \emph{option 1}, the calculator is used to
calculate the photon-noise limited RV measurement precision \sigRV{} using the formalism discussed in
Sect.~\ref{sect:sigrv} and PHOENIX stellar models. Two suboptions are available for
users to either add additional RV noise sources which are sampled from appropriate empirical distributions
(\emph{option 1.1}) or for the user to specify verbatim those additional noise sources
(\emph{option 1.2}). In \emph{option 2}, the user can input a fixed value of \sigRV{} thus
negating the need for certain input parameters to be specified by the user and
speeding up the wall time of the calculation. \emph{Option 2} also features the two suboptions available
for \emph{option 1} and additionally has a third option in which the effective RV rms 
(i.e. the combination of all RV noise sources) is set verbatim if its value is known for the system of
interest and only white noise calculations are desired (\emph{option 2.3}).
\emph{Option 2} may be viable for users whose employed
spectrograph features an independent ETC, the results from which differ from those returned by
the built-in \texttt{RVFC} ETC. 
The input parameters required to run the \texttt{RVFC} are summarized in Table~\ref{table:rvfc}.

\begin{deluxetable*}{lcccc}
\tabletypesize{\scriptsize}
\tablecaption{Descriptions of \texttt{RVFC} Input Parameters \label{table:rvfc}}
\tablewidth{0pt}
\tablehead{Parameter & Units & Required for & Usage & Notes \\ & & \texttt{RVFC} Option(s) & &}
\startdata
\emph{Spectrograph parameters} & & & \\
Minimum spectrograph wavelength & nm & 1 & Interpolating \sigRV{.} & \\
Maximum spectrograph wavelength & nm & 1 & Interpolating \sigRV{.} & \\
Spectral resolution, $R$ & $\lambda / \Delta \lambda$ & 1 & Interpolating \& calculating \sigRV{.} & \\
Telescope aperture & m & 1 & Calculating \sigRV{.} & \\
Throughput & [0-1] & 1 & Calculating \sigRV{.} & \\
RV noise floor, $\sigma_{\text{floor}}$ & \mps{} & 1 & Contributes to RV rms. & \\
\hline \smallskip

\emph{Planet parameters} & & & \\
Orbital period, $P$ & days & 1,2 & Calculating the expected $K$ and & \\ & & & sampling unseen planets. & \\
Planetary radius, $r_p$ & R$_{\oplus}$ & 1,2 & Estimating $m_p$ from Eq.~\ref{eq:MR} and & \\ &&& sampling unseen planets. & \\ 
Planetary mass, $m_p$ & M$_{\oplus}$ & 1,2 & Calculating the expected $K$. & If unspecified, $m_p$ is estimated from \\ &&&& $r_p$ and the mass-radius relation (Eq.~\ref{eq:MR}). \\
\hline \smallskip

\emph{Stellar parameters} & & & \\
Apparent magnitude, $V$ or $J$ & - & 1 & Calculating \sigRV{.} & The spectral coverage of the \\ &&&& spectrograph must span either $V$ or $J$. \\
Stellar mass, $M_s$ & M$_{\odot}$ & 1,2 & Calculating the expected $K$, & \\ &&& $\log{g}$ for \sigRV{,} & \\ &&& sampling stellar activity, and & \\ &&& sampling unseen planets. & \\
Stellar radius, $R_s$ & R$_{\odot}$ & 1,2 & Calculating $\log{g}$ for \sigRV{.} & \\ 
Effective temperature, $T_{\text{eff}}$ & K & 1,2 & Interpolating \sigRV{,} & \\ &&& estimating stellar colors, & \\ &&& sampling stellar activity, and & \\ &&& sampling unseen planets. & \\ 
Metallicity, $Z$ & [Fe/H] & 1,2 & Interpolating \sigRV{.} & \\
Projected rotation velocity, \vsini{} & km s$^{-1}$ & 1 & Interpolating \sigRV{.} & \\
Rotation period, \prot{} & days & 1,2 & Calculating \sigRV{} and & If unspecified, \prot{} is estimated from \\ &&& sampling stellar activity & $R_s$ and \vsini{} assuming $\sin{i_s}=1$. \\
\hline \smallskip

\emph{RV noise sources} & & & \\
Photon-noise limited RV precision, \sigRV{} & \mps{} & 2 & Contributes to RV rms. & \\
RV activity rms, \sigact{} & \mps{} & 1,2 & Contributes to RV rms. & Can be set to zero or is sampled \\ &&&& if left unspecified. \\
RV rms from unseen planets, \sigplan{} & \mps{} & 1,2 & Contributes to RV rms. & Can be set to zero or is sampled \\ &&&& if left unspecified. \\
Effective RV rms, \sigeff{} & \mps{} & 2 & Calculating \nrv{.} & If unspecified, \sigeff{} is computed \\ &&&& from above contributing noise sources. \\
\hline \smallskip

\emph{Simulation parameters} & & & \\
Exposure time, $t_{\text{exp}}$ & minutes & 1,2 & Calculating S/N and the & \\ &&& total observing time. & \\
Overhead time & minutes & 1,2 & Calculating \sigRV{.} & \\
Desired $K$ detection significance, $K/\sigma_{\text{K}}$ & - & 1,2 & Calculating \nrv{.} & \\
Number of GP trials, $N_{\text{GP}}$ & - & 1,2 & Calculating \nrv{} in the & If zero, only do the white noise \\ &&& presence of red noise. & calculation. If $N_{\text{GP}}>0$ then we \\ &&&& recommend setting $N_{\text{GP}} \gtrsim 10$ for \\ &&&& decent sampling. Value returned \\ &&&& is the median of all trials if $N_{\text{GP}}\geq 2$.
\enddata
\end{deluxetable*}

One notable bottleneck in the wall-time of running the \texttt{RVFC} is the time required to compute
the photon-noise limited RV precision given a unique set of stellar and spectrograph parameters. To
facilitate \emph{rapid} calculations  with the \texttt{RVFC} we opt to interpolate these values from
pre-computed tables rather than perform the calculations explicitly. The tables from which \sigRV{}
values are interpolated from are computed individually
for each of the spectral bands shown in Table~\ref{table:bands} and
over five additional parameters: the spectral resolution, $T_{\text{eff}}$, $\log{g}$, $Z$, and \vsini{.}
Given values for these parameters in the \texttt{RVFC}, the corresponding \sigRV{} is obtained by
interpolating over this grid for each spectral band spanned by the spectrograph's wavelength domain. 
The remaining spectrograph parameters and stellar magnitude are then used to scale the interpolated
value of \sigRV{} to the correct S/N per resolution element. Notably, 
the interpolation of \sigRV{} necessitates a trade-off between accuracy and computing time. However,
the loss in accuracy we deem acceptable given the often inexact values of the other sources of RV
noise (i.e. instrument stability, activity, and additional unseen planets).

Lastly, recall that the
\texttt{RVFC} can be calculate \nrv{} in either white or correlated RV noise limits according to
our models discussed in Sects.~\ref{sect:fisherwhite} and~\ref{sect:fisherGP} respectively. As noted in
Sect.~\ref{sect:fisherGP}, the results in the latter scenario are dependent on the time-sampling which has
been sampled uniformly in this study over a fixed baseline. This is also adopted in the initial version of
the \texttt{RVFC}. As such, for users interested in calculating \nrv{} in the presence of correlated RV noise
we recommend using multiple calculations (e.g. $N_{\text{GP}} \gtrsim 10$) to obtain the most-likely value and
spread in \nrv{} given a suite of sampled window functions. Users beware that increasing $N_{\text{GP}}$ will
require a correspondingly longer computation time.  
In the future we would like to implement a way for users to upload custom window functions to avoid this
ambiguity.

\section{Fisher information with a quasi-periodic Gaussian process regression model} \label{app:fishergp}
Here we derive the Fisher information matrix terms for a circular keplerian RV model
plus a quasi-periodic GP correlated noise activity model, including an additional scalar jitter parameter.
As discussed in Sect.~\ref{sect:fisherGP}
the keplerian model parameter is solely the RV semi-amplitude $\{K \}$ whilst the GP covariance model has
five hyperparameters $\boldsymbol{\Theta} = \{a, \lambda, \Gamma, P_{\text{GP}}, \sigma_{\text{jitter}} \}$
that describe the quasi-periodic covariance matrix $C$ commonly used when simultaneously fitting RV planets
and stellar activity:

\begin{align}
  k_{ij} &= a^2 \exp{\left[ -\frac{(t_i-t_j)^2}{2 \lambda^2}
      -\Gamma^2 \sin^2{\left(\frac{\pi |t_i-t_j|}{P_{\text{GP}}} \right)} \right]}, \label{appeq:K1} \\
  C_{ij} &= k_{ij} + \delta_{ij} \sigma_{\text{RV},i}^2. \label{appeq:K2}
\end{align}
  
\noindent We therefore
have six model parameters leading to a $6 \times 6$ Fisher information matrix $B$ which is related
the model parameter covariance matrix $C'=B^{-1}$ from which model parameter measurement uncertainties
are calculated. Assuming a circular keplerian orbit for the transiting planet of interest, the residual
RV vector is

\begin{equation}
  \mathbf{r}(\mathbf{t}) = \mathbf{y}(\mathbf{t}) - (-K \sin{(\phi(\mathbf{t}))})
  \label{appeq:residual}
\end{equation}

\noindent where $\mathbf{y}$ are the raw RVs observed at times $\mathbf{t}$ and $\phi(\mathbf{t})$ is
the planet's orbital phase centered on mid-transit. The generalized lnlikelihood from which the Fisher
information matrix is calculated is then

\begin{equation}
  \ln{\mathcal{L}} = -\frac{1}{2} \left(\mathbf{r}^{\text{T}} C^{-1} \mathbf{r} + \ln{\text{det}C} +
  \text{constant} \right), 
  \label{appeq:lnl}
\end{equation}

Before populating the Fisher information matrix, we note two crucial mathematical identities

\begin{align}
  \frac{\partial}{\partial \theta} C^{-1} &= -C^{-1} \frac{\partial C}{\partial \theta} C^{-1}, \\
  \frac{\partial}{\partial \theta} \ln{\text{det}C} &= \text{tr} \left(C^{-1} \frac{\partial C}{\partial \theta} \right).
\end{align}

We can now proceed with calculating the general equation for each of the 21 unique Fisher information matrix
entries in terms of partial derivatives of either the RV residual vector or covariance matrix $C$ with respect to
the model parameters instead of the partial derivative of the inverse covariance matrix. 

\begin{align}
  B_{ij} &= -\frac{\partial^2 \ln{\mathcal{L}}}{\partial \theta_i \partial \theta_j}, \\
  \frac{\partial \ln{\mathcal{L}}}{\partial \theta_i} &= \frac{\partial}{\partial \theta_i} \left[
    -\frac{1}{2} \mathbf{r}^{\text{T}} C^{-1} \mathbf{r} -\frac{1}{2} \ln{\text{det}C} \right] \notag \\
  &= -\frac{1}{2} \left[ \left(\frac{\partial \mathbf{r}}{\partial \theta_i} \right)^{\text{T}} C^{-1} \mathbf{r}
    - \mathbf{r}^{\text{T}} C^{-1} \frac{\partial C}{\partial \theta_i} C^{-1} \mathbf{r} 
    + \mathbf{r}^{\text{T}} C^{-1} \left( \frac{\partial \mathbf{r}}{\partial \theta_i} \right)
    + \text{tr}\left( C^{-1} \frac{\partial C}{\partial \theta_i} \right) \right], \\
  \frac{\partial^2 \ln{\mathcal{L}}}{\partial \theta_i \partial \theta_j} &= -\frac{1}{2} \Bigg[ \Bigg.
    \left(\frac{\partial^2 \mathbf{r}}{\partial \theta_i \partial \theta_j} \right)^{\text{T}} C^{-1} \mathbf{r}
    - \left(\frac{\partial \mathbf{r}}{\partial \theta_i} \right)^{\text{T}} C^{-1} \frac{\partial C}{\partial \theta_j} C^{-1} \mathbf{r}
    + \left(\frac{\partial \mathbf{r}}{\partial \theta_i} \right)^{\text{T}} C^{-1} \left( \frac{\partial \mathbf{r}}{\partial \theta_j} \right) \notag \\
    &- \Bigg( \Bigg. \left(\frac{\partial \mathbf{r}}{\partial \theta_j} \right)^{\text{T}} C^{-1} \frac{\partial C}{\partial \theta_i} C^{-1} \mathbf{r} 
    - \mathbf{r}^{\text{T}} C^{-1} \frac{\partial C}{\partial \theta_j} C^{-1} \frac{\partial C}{\partial \theta_i} C^{-1} \mathbf{r}
    + \mathbf{r}^{\text{T}} C^{-1} \frac{\partial^2 C}{\partial \theta_i \partial \theta_j} C^{-1} \mathbf{r} \notag \\
    &- \mathbf{r}^{\text{T}} C^{-1} \frac{\partial C}{\partial \theta_i} C^{-1} \frac{\partial C}{\partial \theta_j} C^{-1} \mathbf{r}
    + \mathbf{r}^{\text{T}} C^{-1} \frac{\partial C}{\partial \theta_i} C^{-1} \left( \frac{\partial \mathbf{r}}{\partial \theta_j} \right) \Bigg. \Bigg) \notag \\
    &+ \left( \frac{\partial \mathbf{r}}{\partial \theta_j} \right)^{\text{T}} C^{-1} \frac{\partial \mathbf{r}}{\partial \theta_i}
    - \mathbf{r}^{\text{T}} C^{-1} \frac{\partial C}{\partial \theta_j} C^{-1} \frac{\partial \mathbf{r}}{\partial \theta_i}
    + \mathbf{r}^{\text{T}} C^{-1} \frac{\partial^2 \mathbf{r}}{\partial \theta_i \partial \theta_j} \Bigg. \Bigg] \label{appeq:2nd}
\end{align}

Each matrix entry is calculated using Eq.~\ref{appeq:2nd}. In Eq.~\ref{appeq:2nd} there are
two first order and two second order partial derivatives that must be computed with respect to each of the six model parameters:
$\boldsymbol{\Theta} = \{K, a, \lambda, \Gamma, P_{\text{GP}}, \sigma_{\text{jitter}} \}$. These being

\begin{equation}
  \frac{\partial \mathbf{r}}{\partial \theta_i}, \frac{\partial K}{\partial \theta_i},
  \frac{\partial^2 \mathbf{r}}{\partial \theta_i \partial \theta_j}, \text{ and } \frac{\partial^2 K}{\partial \theta_i \partial \theta_j}  
\end{equation}
  
\noindent and can be computed analytically or symbolically using the open-source \texttt{sympy} package in \texttt{python} given the
analytical expressions for the residual vector (Eq.~\ref{appeq:residual}) and the covariance matrix (Eq1.~\ref{appeq:K1} \&~\ref{appeq:K2}).

\bibliographystyle{apj}
\bibliography{refs}

\clearpage
\begin{turnpage}
\begin{deluxetable}{cccccccccccccccccc}
\tabletypesize{\scriptsize}
\tablecaption{Stellar parameters from the \citetalias{sullivan15} synthetic catalog \label{table:stars}}
\tablewidth{0pt}
\tablehead{TOI & $\alpha$ & $\delta$ & $P$ & $m_p$ & $K$ & $S$ & $M_s$ & $T_{\text{eff}}$ & Distance & $B$ & $V$ & $R$ & $Y$ & $J$ & $H$ & $\text{med}(v\sin{i_s})$ & \\ & $[$deg$]$ & $[$deg$]$ & $[$days$]$ & $[\text{M}_{\oplus}]$ & $[$m s$^{-1}]$ & $[\text{S}_{\oplus}]$ & $[\text{M}_{\odot}]$ & $[$K$]$ & $[$pc$]$ & & & & & & & $[$km s$^{-1}]$}
0000 & 0.44 & 45.22 & 9.135 & 8.19 & 2.03 & 361.7 & 1.40 & 6531 & 100.0 & 8.87 & 8.47 & 8.24 & 7.80 & 7.63 & 7.46 & 2.17 \\ 
0001 & 0.48 & -66.20 & 14.200 & 5.57 & 3.11 & 2.1 & 0.33 & 3426 & 60.3 & 16.52 & 15.08 & 14.17 & 12.05 & 11.56 & 10.96 & 0.23 \\ 
0002 & 0.65 & 42.94 & 4.957 & 4.51 & 1.66 & 235.0 & 1.05 & 5546 & 97.7 & 10.86 & 10.12 & 9.72 & 9.09 & 8.81 & 8.46 & 3.24 \\ 
0003 & 0.92 & -26.07 & 2.160 & 4.32 & 1.95 & 1240.0 & 1.17 & 5984 & 53.7 & 8.61 & 8.06 & 7.76 & 7.20 & 6.98 & 6.73 & 5.81 \\ 
0004 & 1.31 & -24.95 & 9.753 & 5.81 & 2.95 & 5.9 & 0.46 & 3622 & 66.1 & 15.59 & 14.19 & 13.32 & 11.47 & 10.99 & 10.34 & 0.57 \\ 
0005 & 1.38 & 10.61 & 13.990 & 5.89 & 3.32 & 2.1 & 0.33 & 3425 & 58.9 & 16.48 & 15.04 & 14.13 & 12.01 & 11.52 & 10.92 & 0.18 \\ 
0006 & 1.78 & -71.93 & 8.420 & 8.15 & 5.23 & 4.5 & 0.35 & 3444 & 70.8 & 16.73 & 15.29 & 14.38 & 12.28 & 11.79 & 11.19 & 0.27 \\ 
0007 & 1.79 & -9.14 & 5.618 & 7.03 & 9.15 & 2.6 & 0.15 & 3228 & 23.4 & 16.63 & 15.05 & 13.98 & 11.66 & 11.13 & 10.63 & 0.09 \\ 
0008 & 1.95 & -17.00 & 1.344 & 37.81 & 16.03 & 10160.0 & 1.63 & 6668 & 100.0 & 7.78 & 7.51 & 7.37 & 6.83 & 6.72 & 6.63 & 8.19 \\ 
0009 & 2.17 & -15.53 & 17.140 & 11.57 & 2.10 & 341.2 & 1.63 & 6668 & 100.0 & 7.78 & 7.51 & 7.37 & 6.83 & 6.72 & 6.63 & 7.54 \\ 
\tablecomments{Only a portion of this table is shown here to demonstrate its form and content. A machine-readable version of the full table is available.}
\end{deluxetable}
\end{turnpage}
\clearpage
\global\pdfpageattr\expandafter{\the\pdfpageattr/Rotate 90}
\clearpage
\begin{turnpage}
\begin{deluxetable}{ccccccccccc}
\tabletypesize{\scriptsize}
\tablecaption{Median radial velocity noise sources and follow-up calculations for $3\sigma$ planet mass detections of the \citetalias{sullivan15} synthetic catalog \label{table:results}}
\tablewidth{0pt}
\tablehead{TOI & $ \sigma_{\text{RV},\text{opt}} $ & $ \sigma_{\text{RV},\text{nIR}} $ & $ \sigma_{\text{act}} $ & $ \sigma_{\text{planets}} $ & $ \sigma_{\text{eff,opt}} $ & $ \sigma_{\text{eff,nIR}} $ & $ N_{\text{RV,opt}} $ & $ N_{\text{RV,nIR}} $ & $ t_{\text{obs,opt}} $ & $ t_{\text{obs,nIR}} $ \\ & $[$m s$^{-1}]$ & $[$m s$^{-1}]$ & $[$m s$^{-1}]$ & $[$m s$^{-1}]$ & $[$m s$^{-1}]$ & $[$m s$^{-1}]$ & & & [nights] & [nights]}
0000 & 1.02 & 3.14 & 1.25 & 0.00 & 1.61 & 3.59 & 14.8 & 67.6 & 0.4 & 1.6 \\ 
0001 & 10.18 & 8.21 & 7.84 & 2.00 & 13.06 & 11.62 & 294.1 & 159.1 & 42.0 & 3.8 \\ 
0002 & 2.30 & 3.74 & 0.97 & 0.00 & 2.41 & 3.96 & 57.8 & 155.7 & 1.4 & 3.7 \\ 
0003 & 1.09 & 2.77 & 15.11 & 0.00 & 15.16 & 15.41 & 87.2 & 125.7 & 2.1 & 3.0 \\ 
0004 & 4.15 & 6.06 & 8.15 & 0.77 & 9.22 & 10.33 & 66.9 & 115.2 & 8.7 & 2.7 \\ 
0005 & 9.94 & 8.04 & 7.56 & 1.18 & 12.62 & 11.17 & 241.4 & 111.2 & 34.5 & 2.6 \\ 
0006 & 11.55 & 9.33 & 7.64 & 1.21 & 14.40 & 12.60 & 128.4 & 69.8 & 18.3 & 1.7 \\ 
0007 & 8.99 & 6.69 & 7.93 & 3.37 & 12.86 & 11.30 & 34.6 & 32.0 & 4.9 & 0.8 \\ 
0008 & 0.97 & 4.43 & 0.10 & 0.00 & 2.11 & 5.47 & 19.8 & 20.0 & 0.5 & 0.5 \\ 
0009 & 0.91 & 4.11 & 0.76 & 0.00 & 1.78 & 4.85 & 14.8 & 108.5 & 0.4 & 2.6 \\ 
\tablecomments{The resulting number of RV measurements and total observing times reported here are computed in the general case of RVs in the presence of correlated noise (see Sect.~\ref{sect:fisherGP}). Only a portion of this table is shown here to demonstrate its form and content. A machine-readable version of the full table is available.}
\end{deluxetable}
\end{turnpage}
\clearpage
\global\pdfpageattr\expandafter{\the\pdfpageattr/Rotate 90}

\end{document}